\documentclass{sig-alternate}

\newif\ifsubmission
\submissionfalse %

\usepackage{times}
\usepackage{graphicx}
\usepackage{amssymb}
\usepackage{amsmath}
\usepackage{colortbl}
\usepackage{color}
\usepackage{multirow}
\usepackage{threeparttable}
\usepackage{xspace}
\usepackage[table]{xcolor}
\usepackage{flushend}
\usepackage{paralist}
\usepackage[small,bf]{caption}
\usepackage[labelformat=simple]{subcaption}

\usepackage{rotating}
\usepackage{placeins}
\usepackage{mathtools}

\usepackage{appendix}

\def\P{\textsf{P}}

\def\S{\ensuremath{S}}

\definecolor{darkgreen}{RGB}{47,109,79}
\definecolor{darkblue}{RGB}{57,79,99}
\usepackage[bookmarks=false, colorlinks=true, plainpages = false,  linkcolor = darkgreen,   citecolor = darkgreen, urlcolor = darkblue, filecolor = blue]{hyperref}

\usepackage{indentfirst}
\newcommand{\descr}[1]{\medskip\noindent \textbf{#1}}
\usepackage{enumitem}

\newcommand{\server}{\textsf{S}\xspace} 
\newcommand{\user}{\textsf{U}\xspace} 
\newcommand{\userA}{\textsf{A}\xspace} 
\newcommand{\userB}{\textsf{B}\xspace} 
\newcommand{\userC}{\textsf{C}\xspace} 
\newcommand{\userP}{\textsf{P}\xspace} 
\newcommand{\useri}{\textsf{i}\xspace} 
\newcommand{\userj}{\textsf{j}\xspace} 
\newcommand{\ersatzU}{\ensuremath{\mathsf{E}}\xspace}

\newcommand{\remove}[1]{}

\usepackage[hang,flushmargin]{footmisc}

\newcommand{\foffinder}{{\tt Common}~{\tt Friends}\xspace}
\newcommand{\foffinderLine}{{\tt Common Friends}\xspace}
\newcommand{\foffinderttt}{\protect\sectt{Common Friends}\xspace}

\newcommand{\distestLine}{{\tt Social PaL}\xspace}
\newcommand{\distest}{{\tt Social}~{\tt PaL}\xspace}
\newcommand{\distestsect}{{\protect\sectt{Social}~\sectt{Pal}}\xspace}

\newcommand{\sopalconcept}{social path length\xspace}

\ifsubmission

\newcommand{\spotshare}{\texttt{SpotShare}\xspace}
\newcommand{\nearbypeople}{\texttt{nearbyPeople}\xspace}
\else

\newcommand{\spotshare}{\texttt{SpotShare}\xspace}
\newcommand{\nearbypeople}{\texttt{nearbyPeople}\xspace}
\fi

\newcommand{\peershare}{\texttt{PeerShare}\xspace}

\newcommand{\peersharerefonly}{\cite{peershare-paper}\xspace}

\DeclareRobustCommand{\ttfamily}{\fontfamily{lmtt}\selectfont}
\makeatletter
\input{t1lmtt.fd}
\@namedef{T1+lmtt}{}
\makeatother
\DeclareRobustCommand\sectt[1]{{\fontsize{13}{12}\bfseries\ttfamily#1}}

\newcommand{\initiator}{\textsf{A}\xspace}
\newcommand{\responder}{\textsf{B}\xspace}
\newcommand{\puid}{\ensuremath{ID_{\user}}\xspace}    
\newcommand{\pjid}{\ensuremath{ID_{j}}\xspace}

\newcommand{\piid}{\ensuremath{ID_{i}}\xspace}  
\newcommand{\paid}{\ensuremath{ID_{\userA}}\xspace}  
  
\newcommand{\ppid}{\ensuremath{ID_{\userP}}\xspace}
\newcommand{\peid}{\ensuremath{ID_{\ersatzU}}\xspace}    
                     
\newcommand{\cu}{\ensuremath{\mathit{c_{U}}}\xspace} 
\newcommand{\cj}{\ensuremath{\mathit{c_{j}}}\xspace}
\newcommand{\Ci}{\ensuremath{\mathit{C_{i}}}\xspace}
\newcommand{\Fi}{\ensuremath{\mathit{F^{i}}}\xspace}
\newcommand{\Fk}{\ensuremath{\mathit{F^{k}}}\xspace}
\newcommand{\hckj}{\ensuremath{\mathit{c^{k}_{j}}}\xspace}
\newcommand{\hcij}{\ensuremath{\mathit{c^{i}_{j}}}\xspace}
\newcommand{\hciij}{\ensuremath{\mathit{c^{i+1}_{j}}}\xspace}
\newcommand{\hcnj}{\ensuremath{\mathit{c^{n}_{j}}}\xspace}

\newcommand{\Ru}{\ensuremath{\mathit{R_{U}}}\xspace}
\newcommand{\Rr}{\ensuremath{\mathit{R_{B}}}\xspace}
\newcommand{\Ri}{\ensuremath{\mathit{R_{A}}}\xspace}

\newcommand{\Ruh}{\ensuremath{\mathit{R_{U}^{h}}}\xspace}

\newcommand{\Rud}{\ensuremath{\mathit{R_{U}^{d}}}\xspace}
\newcommand{\I}{\ensuremath{\mathit{I}}\xspace}
\newcommand{\Ir}{\ensuremath{\mathit{I_{B}}}\xspace}
\newcommand{\Ii}{\ensuremath{\mathit{I_{A}}}\xspace}

\newcommand{\friendsU}{\ensuremath{\mathit{F(\puid)}}\xspace}

\newcommand{\friendsErsatz}{\ensuremath{\mathit{F(\peid)}}\xspace}
\newcommand{\cE}{\ensuremath{\mathit{c_{\ersatzU}}}\xspace}

\newcommand{\certificate}[1]{\ensuremath{\mathit{Cert_{#1}}}\xspace}           
\newcommand{\sercert}{\certificate{S}}

\newcommand{\update}[1]{#1}

\newcommand{\kPubuserrespDH}{\ensuremath{PK_{B}}\xspace}
\newcommand{\kPubuserinitDH}{\ensuremath{PK_{A}}\xspace}

\newcommand{\bfI}{\ensuremath{BF_{A}}\xspace} %

\newcommand{\kPubInitiator}{\ensuremath{PK_{A}}\xspace}                         %
\newcommand{\kPubResponder}{\ensuremath{PK_{B}}\xspace}                         %

\newcommand{\kSharedIR}{\ensuremath{K_{AB}}\xspace}

\newcommand{\basicserver}{Common Apps Server\xspace}
\newcommand{\appiface}{App Event\xspace}
\newcommand{\dbupdates}{App-DB Updates\xspace}

\newcommand{\mhrw}{MHRW dataset\xspace}

\newcommand{\bfs}{BFS dataset\xspace}

\newcommand{\sampled}{Social Filter dataset\xspace}

\ifsubmission
\newfont{\mycrnotice}{ptmr8t at 7pt}
\newfont{\myconfname}{ptmri8t at 7pt}
\let\crnotice\mycrnotice%
\let\confname\myconfname%

\permission{Permission to make digital or hard copies of all or part of this work for personal or classroom use is granted without fee provided that copies are not made or distributed for profit or commercial advantage and that copies bear this notice and the full citation on the first page. Copyrights for components of this work owned by others than ACM must be honored. Abstracting with credit is permitted. To copy otherwise, or republish, to post on servers or to redistribute to lists, requires prior specific permission and/or a fee. Request permissions from permissions@acm.org.}
\conferenceinfo{WISEC'15,}{June 21--26, 2015, New York, NY, USA}
\copyrightetc{Copyright 2015 ACM \the\acmcopyr}
\crdata{ISBN 978-1-4503-3623-9/15/06\ ...\$15.00.\\
http://dx.doi.org/10.1145/2766498.2766501}

\else
\makeatletter
\let\@copyrightspace\relax
\makeatother
\fi

\title{How Far Removed Are You? Scalable Privacy-Preserving Estimation of Social Path Length with Social PaL\titlenote{\small A preliminary version of this article appears in ACM WiSec 2015. This is the full version.}}

\numberofauthors{6}
\author{
\alignauthor Marcin Nagy\\
                \affaddr{Aalto University} \\
                     \texttt{marcin.nagy@aalto.fi}
\alignauthor Thanh Bui\\
                \affaddr{Aalto University} \\
                      \texttt{thanh.bui@aalto.fi}\\
\alignauthor Emiliano De Cristofaro\\
                \affaddr{University College London} \\
                      \texttt{e.decristofaro@ucl.ac.uk}\\
\and
\alignauthor N. Asokan\\
                \affaddr{Aalto University and \\University of Helsinki} \\
                      \texttt{asokan@acm.org}\\
\alignauthor J\"{o}rg Ott\\
                \affaddr{Aalto University} \\
                      \texttt{jorg.ott@aalto.fi}\\
\alignauthor Ahmad-Reza Sadeghi\\
                \affaddr{TU Darmstadt/CASED} \\
                      \texttt{ahmad.sadeghi@trust.cased.de}\\
}

\ifsubmission
\else
\fi

\begin{document}

\ifsubmission
\else
\pagenumbering{arabic}
\thispagestyle{plain}
\fi

\maketitle

\begin{abstract}
Social relationships are a natural basis on which humans make trust decisions. Online Social Networks (OSNs) are increasingly often used to let users base trust decisions on the existence and the strength of social relationships. While most OSNs allow users to discover the length of the social path to other users, they do so in a centralized way, thus requiring them to rely on the service provider and reveal their interest in each other.

This paper presents \distest, a system supporting the privacy-preserving discovery of arbitrary-length social paths between any two social network users. We overcome the bootstrapping problem encountered in all related prior work, demonstrating that \distest allows its users to find all paths of length two and to  discover a significant fraction of longer paths, even when only a small fraction of OSN users is in the \distest system -- e.g., discovering~70\% of all paths with only 40\% of the users. We implement \distest using a scalable server-side architecture and a modular Android client library, allowing developers to seamlessly integrate it into their apps. 
\end{abstract}

\vspace{-0.2cm}
\section{Introduction}
\label{sec:intro}

The ability to learn the \sopalconcept to other social network users can often help individuals make informed trust and access control decisions.
For instance, if attendees at a large convention
could easily find other attendees with whom they share social links
(e.g., a LinkedIn connection), this could help them decide who to chat or meet up with.
Similarly, travelers and commuters could more consciously decide with whom
to interact, share rides, etc. %
In general, discovering the \sopalconcept between users is beneficial in many interesting scenarios, such as estimating the \emph{familiarity} to  a location (which can in turn be used for context-based security~\cite{MHKSA14}), as well as for routing in delay-tolerant ad-hoc mobile networks~\cite{DH07} and anonymous communications~\cite{johnson2011trust,mittal2012pisces}.

\descr{Problem Statement.} The widespread adoption of Online Social Networks (OSNs)
makes it appealing to %
measure the length of a path between two nodes,
\update{e.g., to use this information as a signal of reciprocal trust and/or social interest.} 
Today, a Facebook user can see the number of common friends
with another user, %
while LinkedIn also displays the \sopalconcept. 
However, as popular OSNs are centralized systems,
so are the features they offer to discover social paths.
\update{As such, they do not particularly adapt well to mobile environments
where social interactions are tied to physical proximity,
thus severely limiting the feasibility of many applications 
scenarios---users may not always be able to connect to the Internet
or willing to reveal their location and/or interests to the provider.
Relying on centralized systems to learn social path lengths implies 
that, whenever Alice queries a server for the
\sopalconcept to Bob, the server learns that Alice is interested in
Bob, their frequency of interactions, and their locations.}
This prompts the need 
for decentralized and privacy-preserving techniques for
\sopalconcept estimation.
Users should only learn if they have any common friends,
without having to reciprocally reveal the identities of friends that they do not share,
and discover the length of the social path between them (for paths longer than two),
without learning which users are in the path.

\descr{Technical Roadmap.} Our work builds on \foffinder~\cite{bfpsi}, a system supporting
privacy-preserving common friend discovery on mobile devices: 
building on a cryptographic primitive called Private Set Intersection (PSI)~\cite{FNP}, it allows mutual friends to be discovered by securely computing the
intersection of friendship
{\em capabilities}, which are periodically distributed from a \foffinder user
to all the friends who are also using it. %
However, besides being limited to social paths of length two,  \foffinder
only discovers the subset of the mutual
friends who are {\em already} in the system, thus suffering from 
an inherent bootstrapping problem.

This paper introduces \distest, the first system that supports the privacy-preserving estimation of the \sopalconcept
between any two social network users.
We introduce the notion of \emph{ersatz nodes}, created for users that are
direct friends of one or more users of \distest but are not
part of the system.  %
We guarantee that two users of \distest will
be able to discover \emph{all} their common friends in the OSN (i.e., all paths of length two).
We then present a hash chain-based protocol that supports
the (private) discovery of social paths longer than two,
and demonstrate its effectiveness by means of extensive simulations.

Our work is not limited to designing protocols: 
we also fully implement \distest and deploy a scalable server architecture
and an Android client library enabling developers
to seamlessly integrate it into their applications. 

\descr{Contributions.} In summary, we make the following contributions:\vspace{-0.15cm}
\begin {enumerate}
\itemsep0em
\item We present an
efficient privacy-preserving estimation
of social paths of arbitrary length (Section~\ref{sec:social}). \vspace{-0.05cm}
\item We state and prove several properties of \distest including the
  fact that, for two users \userA and \userB: (i) \distest will
  find all common friends of \userA and \userB, including those who are not
  using it, and (ii) for each discovered path between \userA and
  \userB, \distest allows each party to compute the \emph{exact} length of the
  path (Section \ref{sec:analysis}). \vspace{-0.05cm}
\item Using samples of the Facebook graph, we empirically show that even when only 40\% of users use the system, \distest will discover more than 70\% of all paths, and over 40\% with just 20\% of the users (Section~\ref{sec:evaluation}). \vspace{-0.05cm}
\item We support Facebook and LinkedIn integration and release the implementation of a scalable server-side architecture and a modular Android client library, allowing developers to easily integrate \distest in their applications  (Section \ref{sec:implementation}). \vspace{-0.05cm}
\item We build two Android apps:
a  friend radar displaying 
the \sopalconcept to  nearby users, and a tethering app enabling users to 
securely share their Internet connection with people with whom they share mutual friends (see Section~\ref{sec:apps}). 
\vspace{-0.1cm}
\end{enumerate}

\section{Background}
\label{sec:background}

\subsection{Private Discovery of Common Friends}\label{subsec:pcd}
We start by discussing the problem of privately discovering common friends, i.e., social paths of length two.
We argue that minimal security properties for this problem %
include both {\em privacy} and {\em authenticity}, as users should neither learn the
identity of non-shared friends nor claim non-existent friendships.

\descr{Private Set Intersection (PSI)~\cite{FNP}.} 
A straightforward approach for privately discovering common friends is to rely on PSI, 
a primitive allowing two parties to learn the intersection of their respective sets 
(and nothing else). If friend lists are encoded as sets, then PSI could be used to privately 
find common friends as the set intersection. One could also
limit disclosure to the {\em number}, but not the identity, of shared friends, using Private Set Intersection 
Cardinality (PSI-CA)~\cite{DGT12}, which only reveals the size of the intersection. 
However, using PSI (or PSI-CA) guarantees privacy but not authenticity, as one cannot prevent users from inserting arbitrary users in their sets and claim non-existent relationships. 

\descr{Bearer Capabilities.} In order to guarantee authenticity, Nagy et al.~\cite{bfpsi} combine bearer capabilities~\cite{DBLP:conf/icdcs/TanenbaumMR86} (aka bearer tokens) with PSI, proposing the \foffinder service, whereby users generate (and periodically refresh) a random number -- the ``capability'' -- and distribute it to their friends via an authentic and confidential channel. As possession of a capability serves as a proof of friendship, users can input it into the PSI/PSI-CA protocol, thereby only learning the identity/number of common and authentic friends.

Since capabilities are large random values, a simpler variant of PSI for private common friend discovery that only relies on cryptographic hash functions and does not involve public-key operations can be used. %
Parties can hash each item in their set and transfer the hash outputs: since the hash is one-way, parties cannot invert it, thus they only learn the set intersection upon finding matching hashes.\footnote{On the other hand, if sets contained low-entropy items, a malicious party could (passively) check whether any item item is in counterpart's set, independently of whether or not it lies in the intersection.}
This can be further optimized using the {\bf\em Bloom Filter based PSI (BFPSI)} primitive (for high-entropy items) outlined in~\cite{bfpsi}. %
On the other hand, it is not clear whether it is possible to do so for PSI-CA, i.e., to only count the number of common friends.

\descr{Bloom Filters.} A Bloom Filter (BF)
is a compact data structure for probabilistic set membership testing~\cite{bloom1970space}.
Specifically, a BF is an array of $\beta$ bits that can be used to represent 
a set of $\alpha$ elements in a space-efficient way.
BFs introduce no false negatives but can have false positives, even though
the probability of a false positive can be estimated (and bounded)
as a function of $\alpha$ and $\beta$.

Formally, let $X=\{x_1, \ldots, x_\alpha\}$ be a set of $\alpha$ elements, and BF be an array of
$\beta$ bits initialized to 0. BF$(j)$ denotes the $j$-th item in BF. 
Then, let $\{h_i: \{0,1\}^* \rightarrow [1, \beta]\}_{i=1}^\gamma$
be $\gamma$ independent cryptographic hash functions,
 salted with random periodically refreshed nonces.
For each element $x \in X$, set $BF(h_i(x))=1$ for $1\leq i \leq \gamma$. To test if $y \in X$, 
check if BF$(h_i(y)) = 1~\forall i$.
An item appears to be in a set even though it was never inserted in the BF
(i.e., a false positive occurs) with probability
$p=(1-(1-1/\beta)^{\gamma\cdot\alpha})^\gamma$.
The optimal size of the filter, for a desired 
probability $p$, can be estimated as:
$\beta=\left\lceil(-\log_{2}{p})/(\ln2)\right\rceil \times \alpha$.

\subsection{\foffinderttt}\label{subsec:system-model}
\begin{figure}[!t]
\centering
\includegraphics[width=0.88\columnwidth]{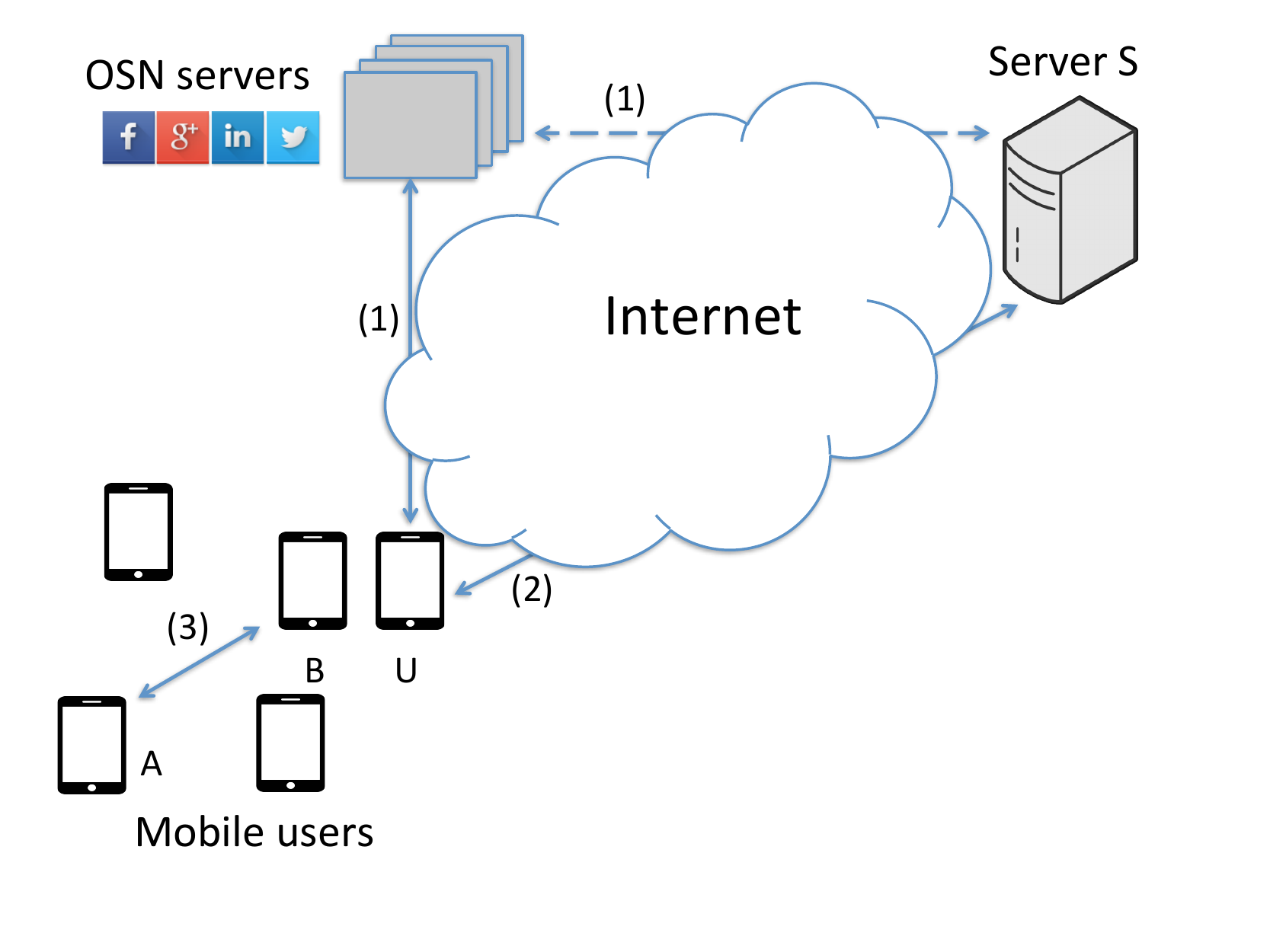}
\vspace{-0.35cm}
\caption{Overview of the \foffinder architecture, involving three
  protocols: (1) OSN user authentication protocol, (2) \foffinder capability distribution protocol, (3) \foffinder discovery protocol.}
\vspace{-0.15cm}
\label{fig:architecture}
\end{figure}

In Figure~\ref{fig:architecture}, we illustrate the \foffinder~\cite{bfpsi} system: it involves a server \server, a set of OSN servers (such as Facebook or LinkedIn), and a set of mobile users, members of one or more of these OSNs. \server is implemented as a social network app 
(i.e., a third-party server), which stores the bearer capabilities uploaded by the \foffinder application running on users' 
devices. It also allows a user's \foffinder application to download bearer capabilities uploaded by that user's friends in the OSNs.
\foffinder consists of three protocols: (1) user authentication,
(2) capability distribution, and (3) common friend discovery. %

\descr{OSN User Authentication} enables the OSN server to
authenticate a user \user, provide \user's OSN identifier
($\puid$) to \server, and \user to authorize \server to access
information about \user's friends in the OSN, which can be done
using standard mechanisms, such as OAuth~\cite{hammer2011oauth}.

\descr{Capability Distribution} 
involves \server and \user, communicating over a
secure channel with server authentication provided by certificate
$\sercert$, and client authentication  based on the previous OSN user
authentication process.
User \user generates a random capability $\cu$ (taken from a large space) and uploads it to \server over the established channel. \server stores $\cu$ along with the social network user identifier $\puid$, and sends back $\Ru=\{(\pjid,\cj):\pjid\in friends(\puid)\}$, which contains pairs of identifiers and corresponding capabilities of each friend that has already uploaded his capability. The protocol needs to be run periodically to keep $\Ru$ up-to-date, as capabilities are periodically refreshed.

\begin{figure}[!t]
\centering
\fbox{\includegraphics[width=0.9\columnwidth]{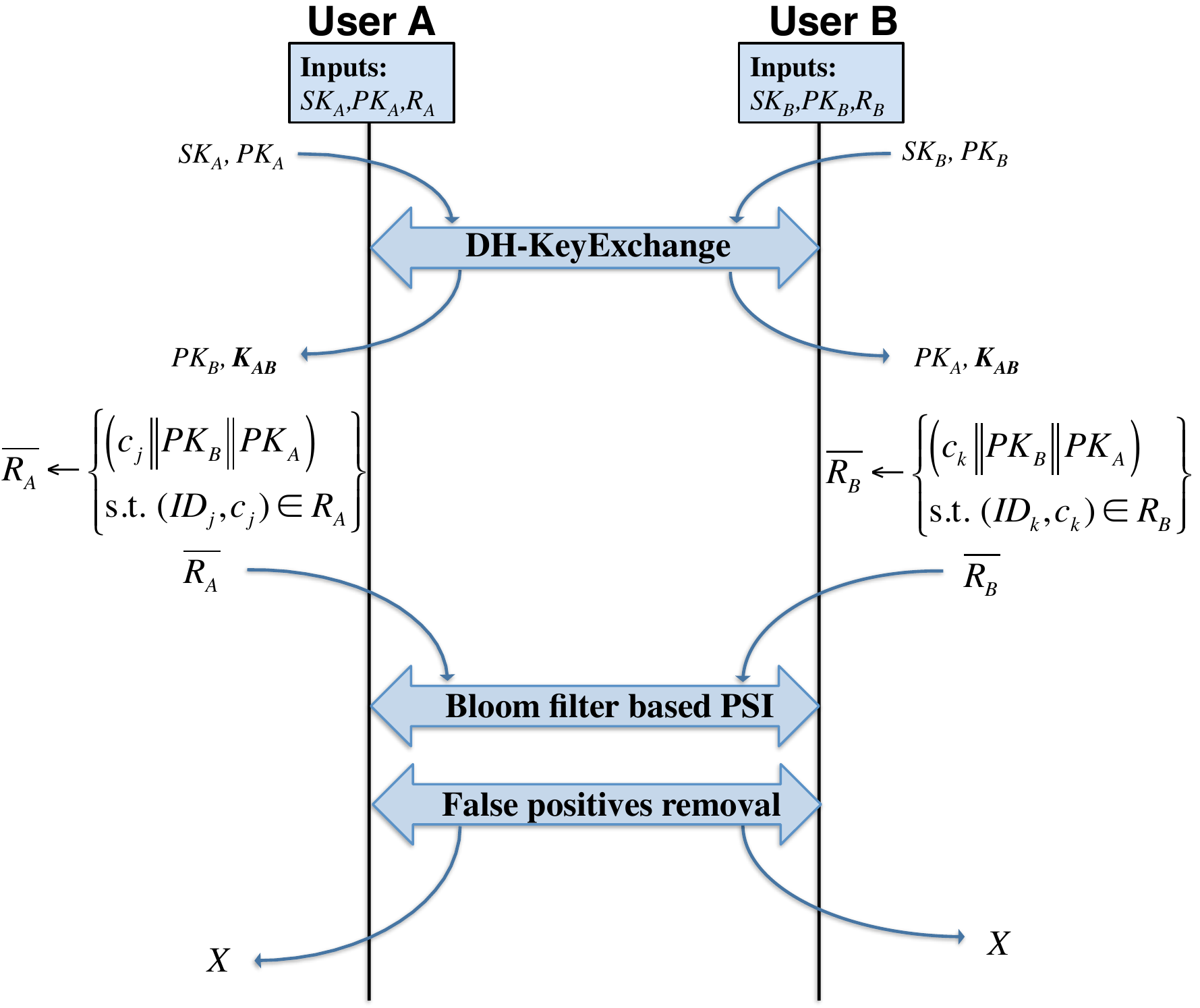}}
\vspace{-0.2cm}
\caption{Common friend discovery protocol based on BFPSI and bearer capabilities from~\cite{bfpsi}.}
\label{fig:BFfof}
\end{figure}

\descr{Common Friend Discovery} is a protocol
run between two users, \initiator and \responder, illustrated it in Figure~\ref{fig:BFfof},
allowing \initiator and \responder to privately discover their common (authentic) friends,
based on BFPSI. 
First, \initiator and \responder exchange their public keys ($\kPubuserinitDH$ and $\kPubuserrespDH$, respectively) and generate a shared key ($\kSharedIR$) used to encrypt messages exchanged as part of the protocol. To prevent man-in-the-middle attacks, \initiator (resp., \responder) cryptographically binds the DH channel to the protocol instance: \initiator (resp., \responder) extends each item in the capability set $\Ri$ (resp., $\Rr$) by appending DH public keys $\kPubuserinitDH$, $\kPubuserrespDH$, building effectively a new set $\overline{\Ri}$ (resp., $\overline{\Rr}$). \initiator inserts every element of the $\overline{\Ri}$ set into a Bloom Filter $\bfI$ and sends it to \responder. \responder discovers the set of friends ($X'$) in common with \initiator by verifying whether each item of his input set $\overline{\Rr}$ is in $\bfI$. Since Bloom Filters introduce false positives, the set $X'$ may contain non-common friends. Thus a simple challenge-response protocol is run, where \responder requires \initiator to prove knowledge of items available in $X'$. At the end of the protocol, \initiator and \responder output the set of their common friends $X$.

\section{Social Pal}
\label{sec:social}

We now present the design and the instantiation of the \distest, the system
to compute the \sopalconcept between two OSN users in a decentralized and privacy-preserving way.

\subsection{System Design}\label{subsec:requirements}

\begin{figure}[!t]
\centering
	\begin{subfigure}[b]{0.35\textwidth}
		\includegraphics[width=0.85\textwidth]{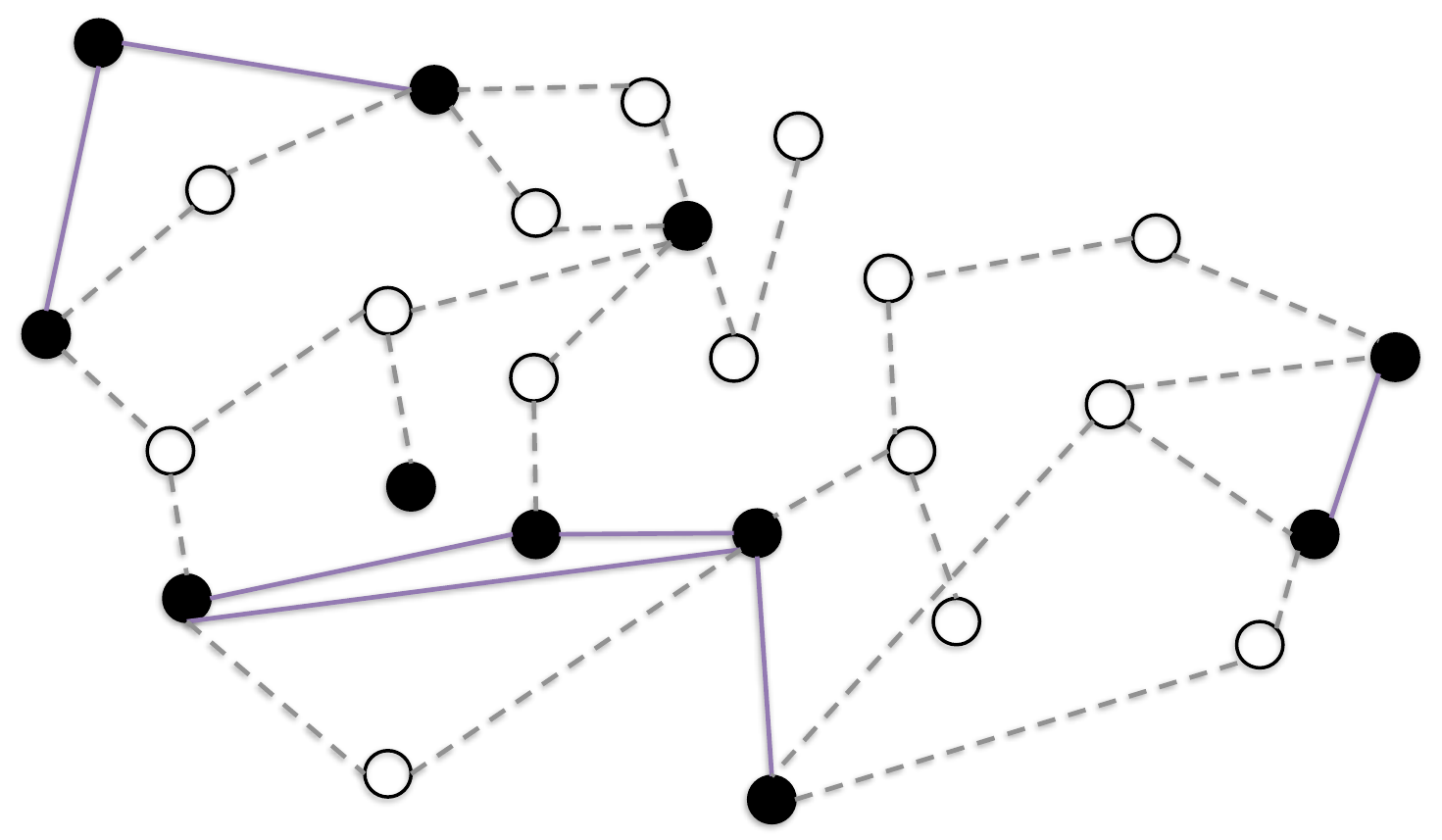}
		\caption{\foffinder, 40\% of active users}
                \label{subfig:0.25-users}
	\end{subfigure}
        \vspace{-0.05cm}
      	\begin{subfigure}[b]{0.35\textwidth}
		\includegraphics[width=0.85\textwidth]{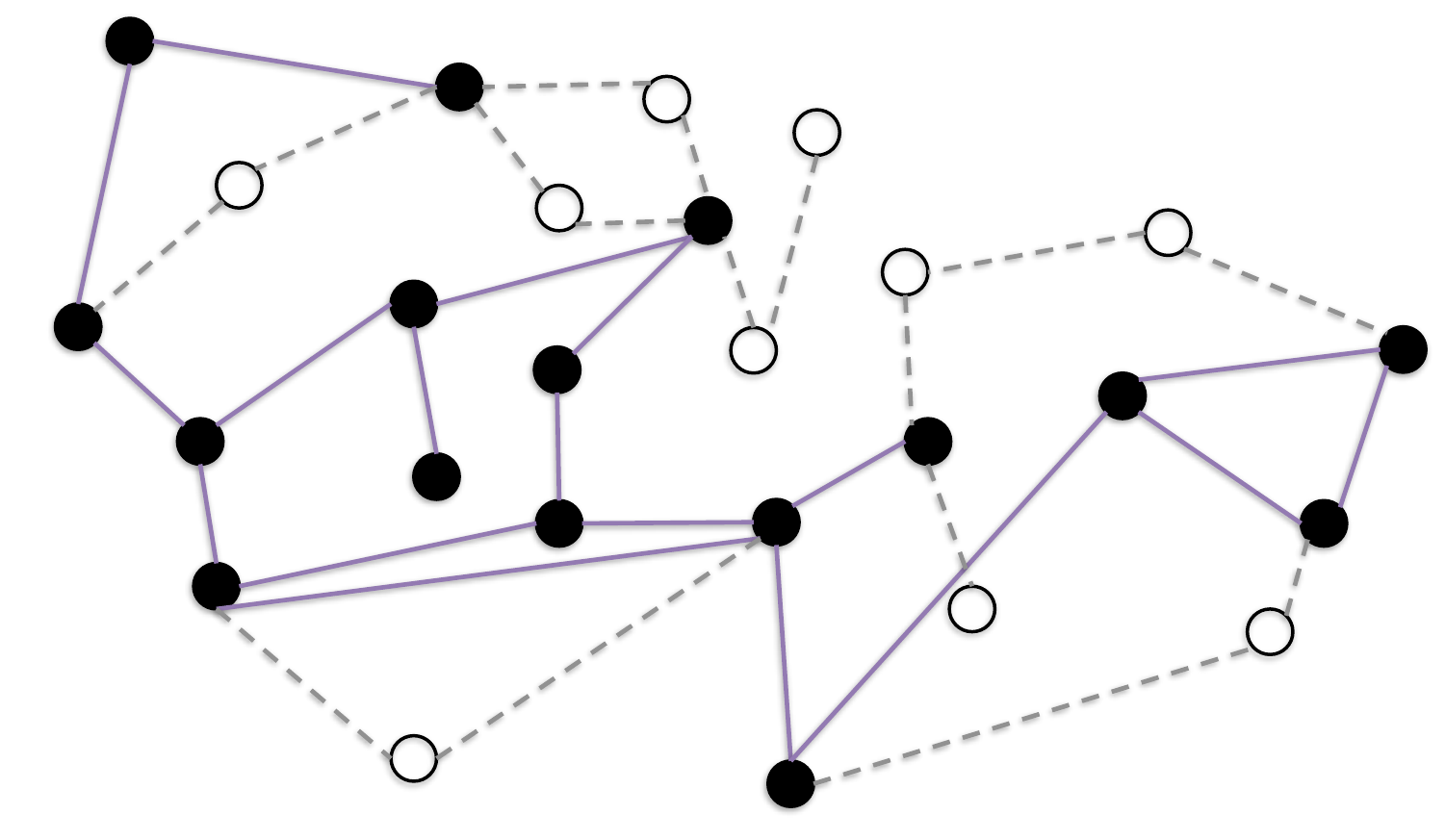}
		\caption{\foffinder, 60\% of active users}
                \label{subfig:0.5-users}
        \end{subfigure}
\caption{Coverage of \foffinder with $40\%$ and $60\%$ of users using the system. Black (resp., white) nodes denote users (resp., non-users) of the system.
Purple/solid edges denote a direct friendship discoverable by \foffinder.}
\label{fig:bootstrapping}
\end{figure}

\descr{Limitations of~\cite{bfpsi}.} Before
introducing \distest's requirements, we discuss two main limitations of \foffinder~\cite{bfpsi}, as
addressing them constitute our starting point:\vspace{-0.1cm}
\begin{enumerate}
\item {\em Bootstrapping}: Users
\userA and \userB can discover a mutual friend (say \userC), only if
\userC has joined the \foffinder system and uploaded his capability to \server.
That is, \foffinder will only discover a subset of the mutual friends
between \userA and \userB until all of them start using it. \vspace{-0.15cm}
\item {\em Longer social paths}: \foffinder only
  allows its users to learn whether they are friends or have mutual friends.
If two users have a longer social path between them, \foffinderLine cannot
detect it. %
\end{enumerate} 
To illustrate \foffinder's bootstrapping problem, we plot, in Figure~\ref{fig:bootstrapping},
a simple social network with 27 nodes (i.e., users) and 34 edges (i.e., friendship relationships).
Black circles represent users who are using \foffinder, and white circles --
those who are not. Purple/solid edges represent direct
friend relationships (i.e., social paths of length 1) that are
discoverable by \foffinder. %
When the user base is only 40\% of all OSN users (Figure~\ref{subfig:0.25-users}), only 7 out of the 34  direct friend relationships are discoverable (i.e., coverage is approximately $20\%$). When it increases to 60\%, %
coverage increases to about 50\% (Figure~\ref{subfig:0.5-users}).

\descr{System model.} \distest's system model is the same as
that of \foffinder (presented in Section~\ref{subsec:system-model}).
It involves a server \server (which we design as a social network app), a set of OSN servers (such as Facebook or LinkedIn), and a set of mobile users members of  one or more of these OSNs (cf. Figure~\ref{fig:architecture}).

\descr{Functional requirements.} Ideally, \distest should 
allow any two users to always compute the exact length of the social path between them,
even %
when \distest is being used only by a fraction of OSN users.
In order to characterize how well \distest meets this
requirement, we use a measure of the likelihood that any two
users would discover, using \distest, an existing social path of a given length between
them. We denote this measure as \distest's {\em coverage}. %

We define \distest's functional requirements as follows:\vspace{-0.1cm}
\begin{itemize}
\item {\em (Correctness).} Users \userA and \userB can determine the exact length of a social path between them (if any).\vspace{-0.1cm}
\item {\em (Coverage Maximization).} \distest should maximize coverage, in other words, the ratio between the number of social paths (of length $n$) between \userA and \userB {\em ~discovered} by \distest and the number of {\em all} social paths (of length $n$) between \userA and \userB.\vspace{-0.1cm}
\end{itemize}

\descr{Privacy requirements.} From a privacy point of view, \distest should satisfy the following
requirements. Let \userA and \userB be two \distest users willing to discover
the length of the social path existing between them:
\begin{enumerate}
\itemsep0cm
\item \userA and \userB discover the set of their common friends but learn nothing about their non-mutual friends;
\item \userA and \userB do not learn any more information other than what it is already available from standard OSN interfaces.
\end{enumerate}
In other words, \distest should allow two users to learn the social path length between them
(if any), but not the nodes on the path, without reciprocally revealing their social link.
If a path between the users exists that is of length two (i.e., users have some common friends),
then they learn the identity of the common friends (and nothing else).
This only pertains to interacting users, as ensuring
that no eavesdropping party learn any information about users' friends can be
achieved by letting users communicate via confidential and authentic channels.

\descr{Threat model.} We assume that the participants in \distest are
honest-but-curious. The OSN server is trusted to correctly
authenticate OSN users and not to attempt posing as any OSN
user. The \distest server \server is trusted to
distribute \distest capabilities only to those \distest users
authorized to receive them. \distest users use the legitimate \distest
client,\footnote{This is enforced by the OSN app interfaces which
  ensure that only designated client apps are allowed to talk to a
  particular OSN app server, i.e., the \distest server \server.} but
they might attempt to learn as much information as possible about
friends of other \distest users with whom they interact. 
We aim to guarantee the privacy requirements discussed above in this setting, and
prevent the OSN server or the server \server to learn any information about
interactions between \distest users.

\subsection{Bootstrapping \distestsect}\label{sec:bootstrapping}
Before presenting the details of the system, in Table~\ref{tab:FoFProtocol-notation}, we introduce some notation used throughout the rest of the paper.

\begin{table}[t]
	\centering
	\resizebox{0.95\columnwidth}{!}{%
	\begin{tabular}{| r | l |}
	\hline
	{\bf Symbol} & {\bf Description} \\ \hline
	\multicolumn{2}{|l|}{\cellcolor{lightgray}\emph{Entities}} \\ \hline
	\server & Server \\ \hline
	\initiator, \responder & User A, B, resp. \\ \hline
	\user & Generic User (can be either \initiator or \responder) \\ \hline
	\ersatzU & User E (ersatz node) \\ \hline
	\multicolumn{2}{|l|}{\cellcolor{lightgray}\emph{Social graph data}} \\ \hline
	\puid & Social identifier of \user \\ \hline
	$F(\puid)$ & Set of direct friends of \user \\ \hline
	$F^{k}(\puid)$ & Set of social contacts $k$ hops from \user  \\ \hline
	\multicolumn{2}{|l|}{\cellcolor{lightgray}\emph{Keys}} \\ \hline
	\kPubInitiator, \kPubResponder & DH public key of \initiator, \responder, resp. \\ \hline
	$\kSharedIR$ & DH session key between \initiator and \responder \\ \hline
	\multicolumn{2}{|l|}{\cellcolor{lightgray}\emph{Cryptographic functions}} \\ \hline
	$h^{i}(x)$ & Hash chain of item $x$ of length $i$\\ \hline
	\multicolumn{2}{|l|}{\cellcolor{lightgray}\emph{\distest protocol data}} \\ \hline
	\cj & Capability uploaded by user with identifier \pjid \\ \hline
	\hckj & Capability of degree $k$ uploaded by user with identifier \pjid \\ \hline
	\Ru & Set of capabilities downloaded by \user from \server \\ \hline
	\Ruh & Set of higher order capabilities downloaded by \user from \server \\ \hline
	\Rud & Set of derived higher order capabilities downloaded by \user from \server \\ \hline
	\I & Union of capabilities' sets \Ru, \Ruh, \Rud \\ \hline
	$\overline{\I}$ & Input sets to \distest discovery protocol \\ \hline
	\bfI & Bloom filter sent by \initiator \\ \hline
	\end{tabular}
	}
	\vspace{-0.2cm}
	\caption{Notation.}
	\vspace{-0.2cm}
	\label{tab:FoFProtocol-notation}
\end{table}

\descr{Ersatz nodes.} One fundamental building block of \distest are ersatz nodes\footnote{The word {\em ersatz}, originally from
German, means ``substitute.''},
which we introduce to overcome the bootstrapping problem faced by services like \foffinder.
Recall from Section~\ref{subsec:system-model}
that, in the original \foffinder design, 
the server \server stores bearer capability $\cu$, uploaded by a user
\user, together with his social network identifier $\puid$. \update{The pair 
$(\puid,\cu)$ constitutes \user's {\em user node} in the social graph maintained by \server. The set of \user's friends $\friendsU$ is the set of edges incident in the user node.} %

\begin{figure}[!t]
	\vspace{0.2cm}
	\centering
	\begin{subfigure}[c]{0.38\textwidth}
	\centering
 		\includegraphics[width=0.83\textwidth]{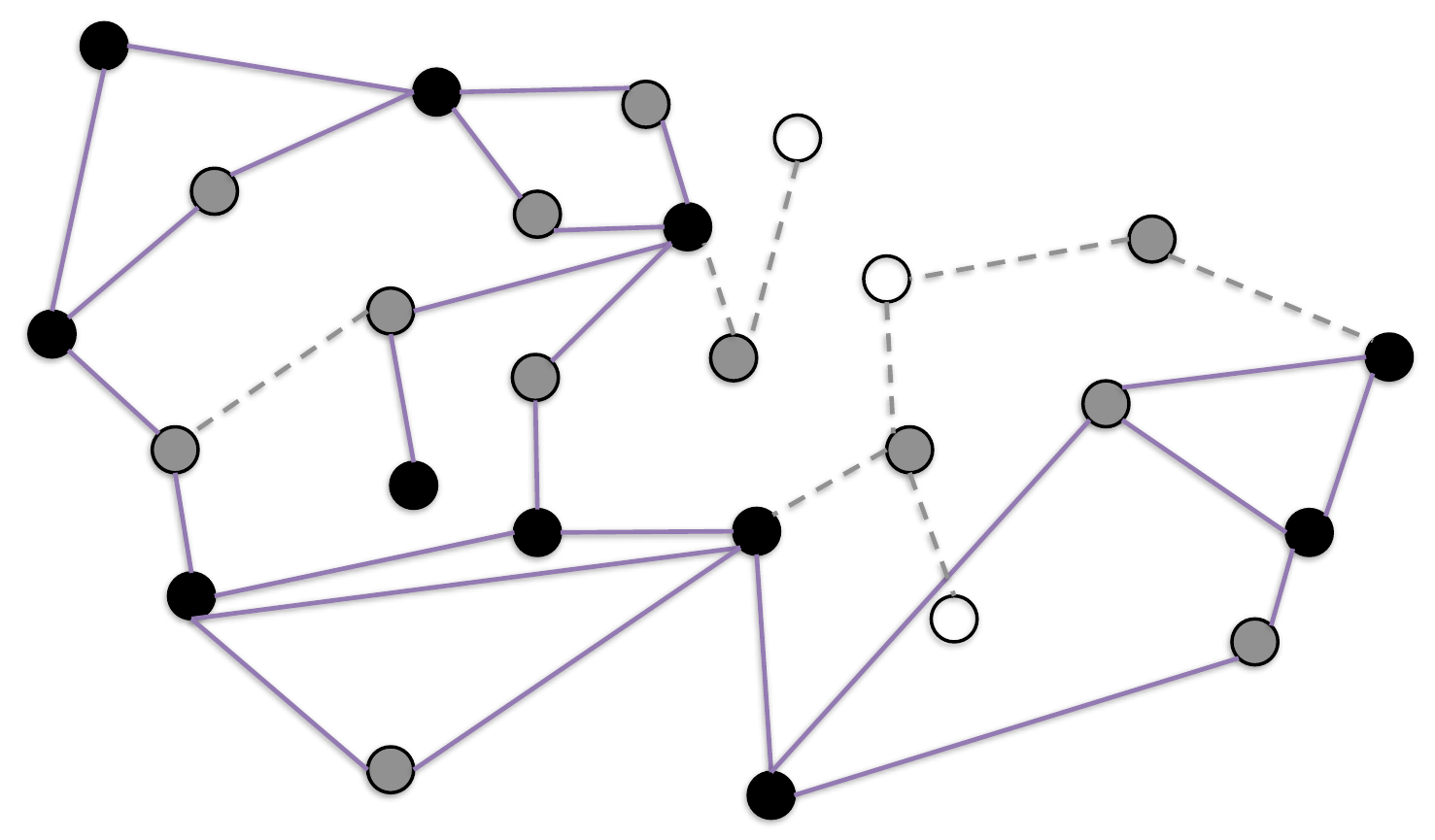}
 		\caption{\distest, 40\% of active users with ersatz nodes}
                 \label{subfig:0.25-users-ersatz}
         \end{subfigure}
 	\begin{subfigure}[c]{0.37\textwidth}
         	\centering
 		\includegraphics[width=0.83\textwidth]{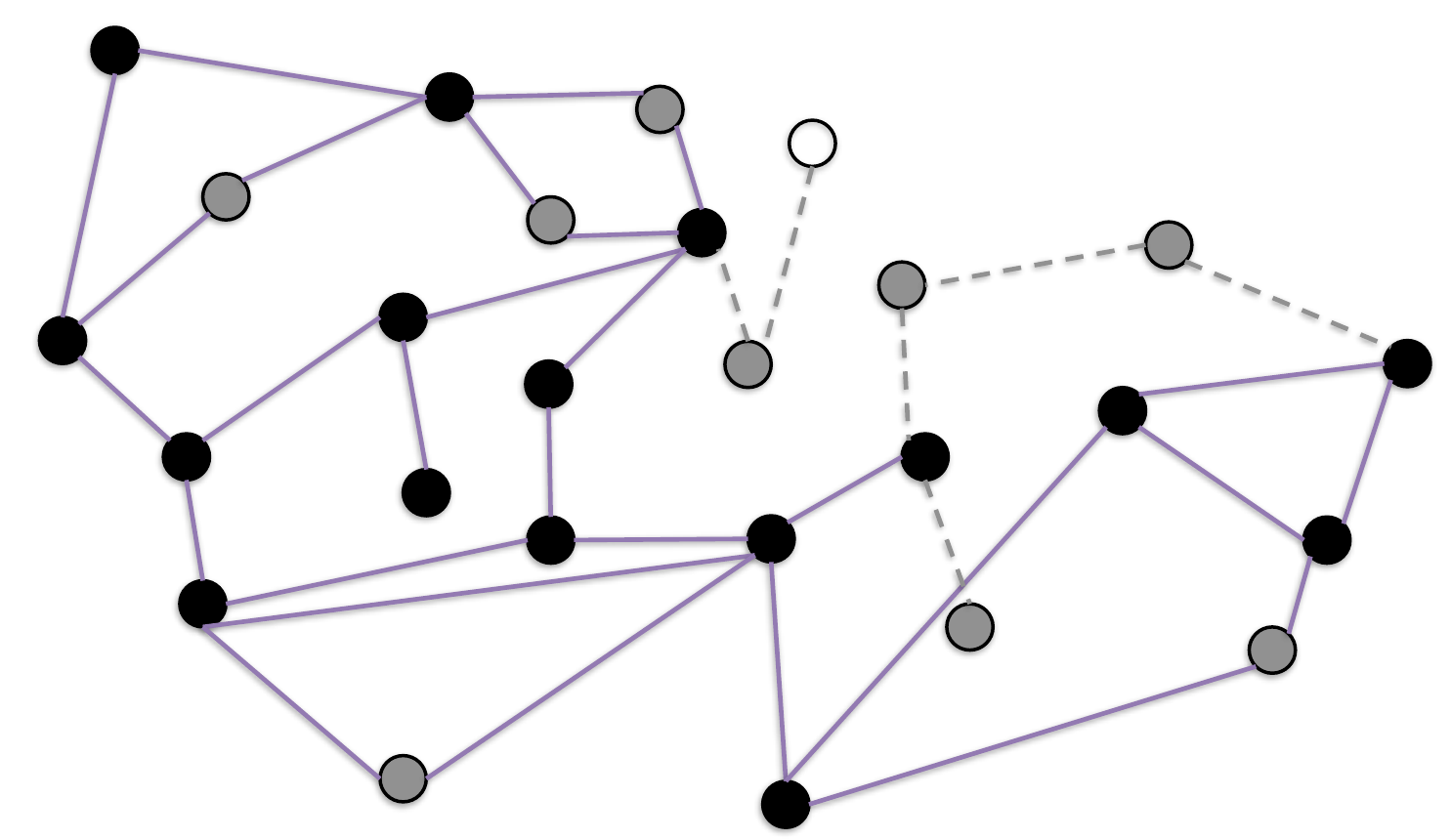}
 		\caption{\distest, 60\% of active users with ersatz nodes}
                 \label{subfig:0.5-users-ersatz}
         \end{subfigure}
         \vspace{-0.2cm}
\caption{Coverage of \distest with $40\%$ and $60\%$ of users using the system
and with the addition of ersatz nodes (in grey).
Purple/solid edges here denote a direct friendship discoverable by \distest.}
         \vspace{0.2cm}

\end{figure}

In \distest, we let \server create an {\em ersatz node} for all users who have not joined the system but who are friends with a user who has. 
An ersatz node is identical to a standard user node, but its capability is generated by \server, 
instead of the user.
Figures~\ref{subfig:0.25-users-ersatz}
and~\ref{subfig:0.5-users-ersatz} show how coverage improves when ersatz nodes are added, 
e.g.,~with only $40\%$ joining the system, coverage reaches $75\%$.

\begin{figure}[ttt]
\centering
\includegraphics[width=1.0\columnwidth]{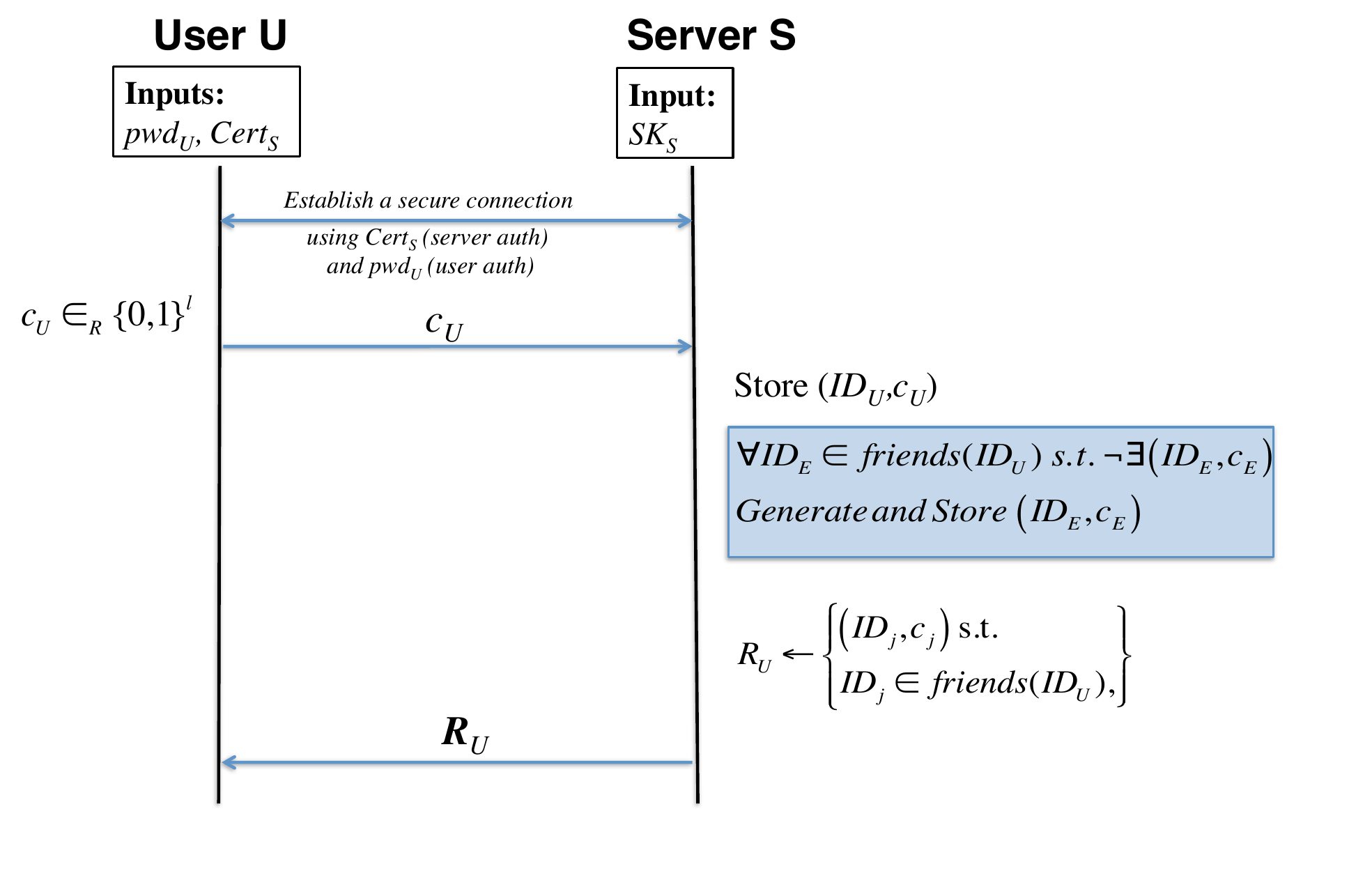}
\caption{Adding ersatz nodes to the capability distribution.}
\vspace{-0.1cm}
\label{fig:ersatz-capability-updates}
\end{figure}

\descr{Ersatz node creation.} 
Adding ersatz nodes requires a few changes in the capability
distribution protocol, compared to that from Section~\ref{subsec:system-model}. We highlight these changes in
Figure~\ref{fig:ersatz-capability-updates}, specifically, in the blue-shaded box.
Before returning $\Ru$ to \user, \server first computes the set
$M_{U}=\{\peid: \neg \exists
(\peid,\cE) \}$ which contains the social network identifier of each
``missing user'' \ersatzU. 

Then, $\forall \peid \in M_{U}$,
it creates \ersatzU's 
an ersatz node as follows:\vspace{-0.1cm}
\begin{enumerate}
\item Create an {\em ersatz capability} $\cE \in _{R} \{0,1\}^{l}$ (where $l$ is the length
  of a capability) for \ersatzU and store
  $\{(\peid,\cE) \}$. \vspace{-0.1cm}
\item Create an initial friend set $\friendsErsatz$, which at this stage contains only $\puid$.\vspace{-0.1cm}
\end{enumerate}
After the successful creation of all needed ersatz nodes, \server
returns $\Ru$, which includes the capabilities from the nodes of all of
\user's friends, including the ersatz nodes.

\descr{Active social graph updates.} 
Users of \distest explicitly authorize the server \server to
retrieve their friend lists from the OSNs. Since an ersatz node
\ersatzU is not a user of \distest, \server cannot learn the full set
of \ersatzU's friends $\friendsErsatz$. Instead, it maintains an
estimate of $\friendsErsatz$ based on the events it can observe from
users of \distest. For example, when a user \user adds \ersatzU as a
friend, \server learns that $\peid$ is added to $\friendsU$ and can infer
that $\puid$ should be added to $\friendsErsatz$.
\update{Each $\puid \in \friendsErsatz$ corresponds to a real user \user who has explicitly authorized \server to learn about the edge \user-\ersatzU in the social graph. %
}

\descr{Turning ersatz nodes into ``standard'' nodes.} 
If a user \ersatzU for whom \server has created an ersatz
node later joins \distest, he
can simply upload his capability $c_{E}$ to \server, who (1) overwrites the old
ersatz capability with $c_{E}$, 
(2) queries the OSN for \ersatzU's friend list
\friendsErsatz, and (3) updates the existing, possibly incomplete, list of \ersatzU's
friends with \friendsErsatz, turning an ersatz node
into a standard node. Note that this operation is transparent to all users.

\subsection{Discovering Longer Social Paths} \label{sec:distance}
We now present the full details of our \distest instantiation: besides addressing the bootstrapping
problem (using ersatz nodes), it also allows two arbitrary users to calculate the \sopalconcept between them.
We denote with $Dist(\userA,\userB)$ the \sopalconcept between two users \userA and \userB, i.e., the minimum number of hops in the social network graph that separates \userA and \userB.

\descr{Intuition.} We set to allow \distest to discover the \sopalconcept between users in the OSN by  extending the capability distribution to include further relationships beyond friendship (e.g., friend-of-a-friend) and rely on capability matches for estimating the \sopalconcept. By using cryptographic hash functions, we can generate and distribute capabilities of higher order that serve as a proof of a social path between users.

\descr{Notation.} In the rest of the paper, we use the following notation:
\begin{compactitem}
\item The hash chain $h^{i}(x)$ of item $x$ (of length $i$) corresponds to the evaluation of a cryptographic hash function $h(\cdot)$ 
performed $i$ times on $x$. When $i=0$, $h(x)=x$. Specifically:
$$h^{i}(x) = \left \{
\begin{array}{l l }
\small
\underbrace{h(h( \cdots ( h}_{i~times}(x))\cdots)) & \quad i\geq 1 \\
x & \quad i=0
\end{array}
\right \} $$
\item \hckj is a $k$-degree capability and is defined as $ \hckj = {h^{k}}(\cj) $
\item $\Fk(\puid),k\geq1$ denotes the set of social contacts that are $k$-hops from user $\user$.\vspace{0.1cm}
\end{compactitem}

\begin{figure}[!t]
\centering
\includegraphics[width=.97\columnwidth]{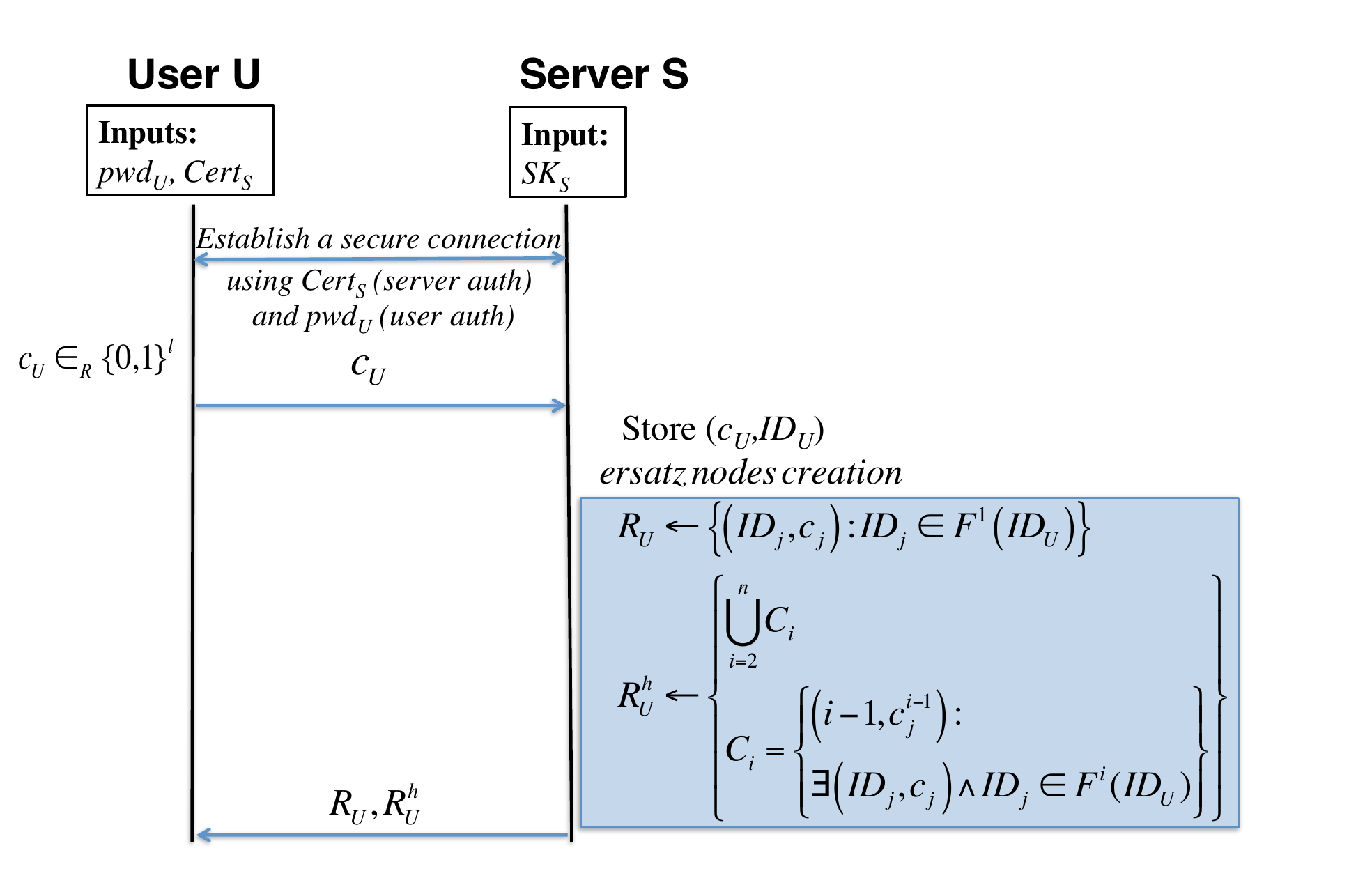}
\caption{\distest's capability distribution protocol.}
\vspace{0.18cm}
\label{fig:distance-capability-protocol}
\end{figure}

\descr{Capability Distribution:}
In Figure~\ref{fig:distance-capability-protocol}, we detail \distestLine's protocol for capability distribution. Interaction between \user and \server is identical to the capability distribution protocol from Figure~\ref{fig:ersatz-capability-updates}, up until the creation of missing ersatz nodes is completed. In the updated protocol \server returns two sets, namely \Ru and \Ruh,
where \Ruh denotes the set of higher order capabilities provided to
\user by other OSN members that are at least $2$-hops from
\user. It is composed of a number of subsets $\Ci, \; i=2,\dots,n$
with each subset \Ci containing $i-1$ order capabilities of users in $\mathit{F^{i}}(\puid)$.  Formally, %
$$
\begin{array}{c}
\Ruh = \bigcup_{i=2}^{n} \Ci, \mbox{and} \\[0.5em]
\Ci=\{ (i-1,c^{i-1}_{j}): \exists (\pjid, \cj) \land \pjid \in \Fi(\puid) \}\vspace{-0.1cm}
\end{array}
$$
Consequently, the total cardinality of \Ruh and \Ru is: \vspace{-0.15cm} $$|\Ru| + |\Ruh| = \sum_{i=1}^{n} |\mathit{F^{i}(\puid)}| \vspace{-0.1cm}$$
Finally, \user generates missing higher order capabilities. For every
received capability \hcij of degree $i$, \user hashes it $n-i$ times
to generate a sequence of higher order capabilities of the form:
$$((i+1,\hciij), \dots, (n,\hcnj))$$
All elements of such sequences are combined into one set of derived higher order capabilities \Rud. Finally all capability sets are combined to form \I:
$$\I=\Ru \cup \Ruh \cup \Rud$$
The resulting set $\I$ will be used to derive the input sets for PSI
during the \sopalconcept discovery protocol as explained below. The
cardinality of the input set to PSI is therefore:\vspace{-0.1cm}
$$| \I |=\sum_{i=1}^{n} |\mathit{F^{i}(\puid)}|\times(n-i+1)\vspace{-0.1cm}$$

To construct $\mathit{F^{n}}(\puid)$ of user
\user, 
\server tracks changes in friend lists of users by using the
following logical implication: 
if $\userj$ represents a friend of $\useri$ 
and $\useri$ is $k-1$ hops from $\user$ 
and $\userj$ was not previously identified at less than
$k$ hops from $\user$,
then $\userj$ is $k$ hops from $\user$. Formally:\vspace{-0.1cm} 
\begin{eqnarray*}
(\piid \in F^{k-1}(\puid) \land \pjid \in F(\piid) \land \\ \pjid
\notin F^{m}(\puid), m < k) \implies \pjid \in F^{k}(\puid)\vspace{-0.1cm}
\end{eqnarray*}

Finally, as capabilities are meant to be short-lived (i.e., they should expire within a couple of days), the protocol needs to be run periodically in order to keep $I$ up-to-date.

\begin{figure}[t]
\centering
\includegraphics[width=0.88\columnwidth]{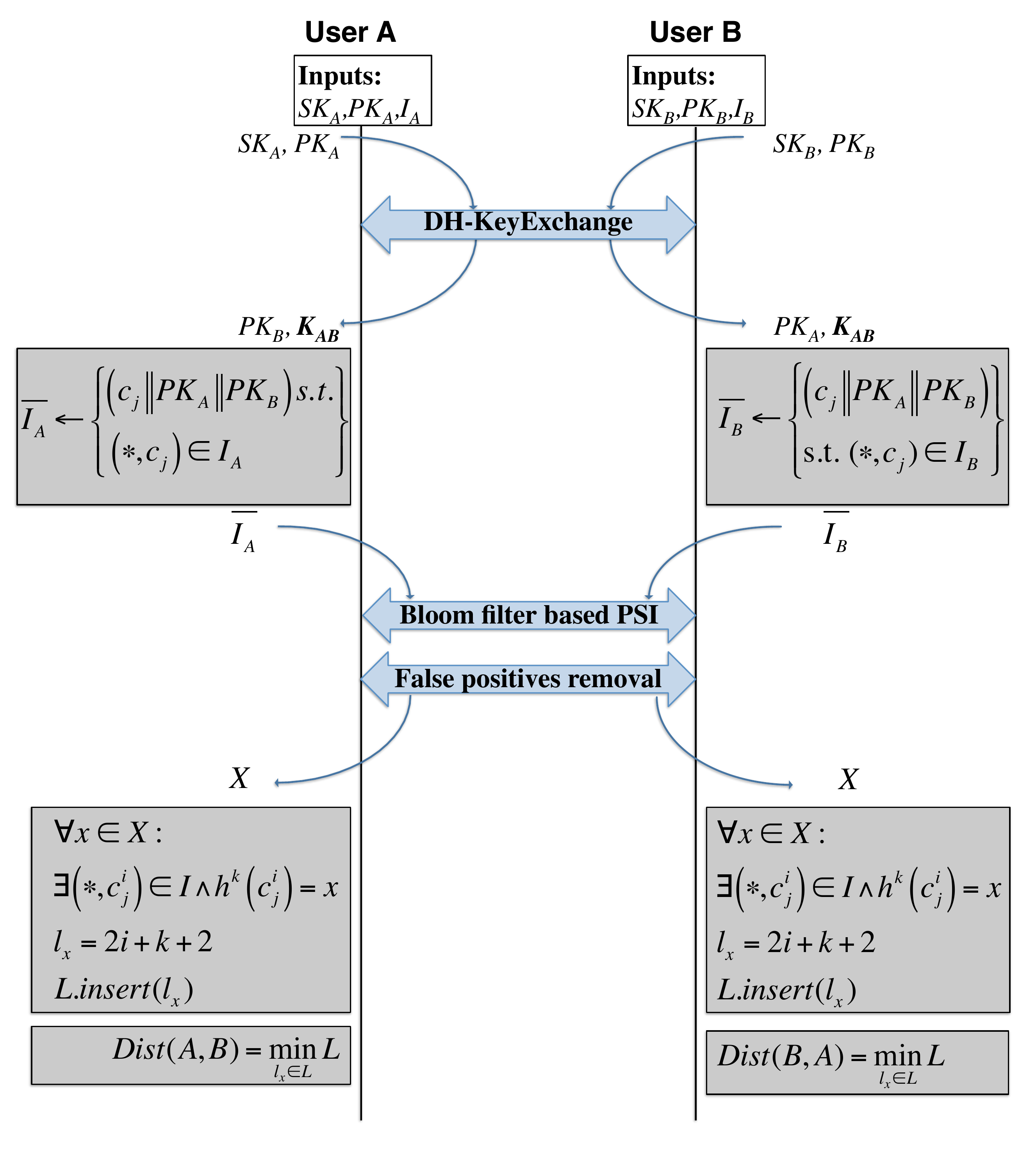}
\vspace{-0.15cm}
\caption{Illustration of \distest discovery protocol. The updated part is marked in grey background.} %
\label{fig:distance-client-protocol}
\end{figure}

\descr{Social Path Discovery:} In
Figure~\ref{fig:distance-client-protocol}, we illustrate the \distest
discovery protocol. 
The protocol involves two users \userA and \userB, who are members of the same OSN. It begins with establishing a secure channel (via Diffie-Hellman key exchange), followed by cryptographic binding of the Diffie-Hellman channel to the protocol instance, which is needed to avoid man-in-the-middle attacks. \userA (resp., \userB) appends both public keys to each capability $\cj$ in $\Ii$ (resp., $\Ir$) set to form $\overline{\Ii}(\overline{\Ir})$. The resulting sets are: \vspace{-0.35cm}

$$\overline{\Ii}=\{ (\cj||\kPubuserinitDH||\kPubuserrespDH): (\ast,\cj) \in \Ii \} \vspace{-0.2cm}$$
$$\overline{\Ir}=\{ (\cj||\kPubuserinitDH||\kPubuserrespDH): (\ast,\cj) \in \Ir \} $$
Note that the $\ast$ symbol in the above equations indicate that, while constructing $\overline{\Ii}$ and $\overline{\Ir}$, the first element of each pair contained in $\Ii$ and $\Ir$ is ignored.

Next, both users execute the steps of \foffinder 's discovery protocol, on the above input sets. Specifically, they 
interact in a Bloom-filter based PSI execution and run the challenge-response part of the protocol needed to remove potential false positives (as discussed in Section~\ref{subsec:system-model}). The interactive protocol ends with parties outputting the intersection of the sets.
From this point on, both users perform identical actions to calculate the social path length between them. All operations are done locally, i.e., with no need to exchange data. This process consists of two phases: (1) calculating the social path length input set $L$ (i.e., the set containing lengths for all discovered paths between \userA and \userB), and (2) selecting  the shortest length among all lengths contained in $L$. To this end, \userA (\userB) builds set $L$ by performing following actions on every item $x \in X$:\vspace{-0.1cm}
\begin{enumerate}
\item Finding a capability $\hcij$ 
such that $\exists (*, \hcij) \in \I \land
 h^k(\hcij) = x$\vspace{-0.1cm}
\item Calculating path length $l_{x}$ via matching capability $x$
  (which was obtained from some user, say \userC) and inserting it into $L$:  \vspace{-0.15cm}
$$
\begin{array}{l}
l_{x}=\underbrace{(i+1)}_{Dist(A,C)}+\underbrace{(i+k+1)}_{Dist(C,B)}=2i+k+2 \\[2ex]
L.insert(l_{x})\vspace{-0.1cm}
\end{array}  
$$
\end{enumerate}
At the end, \userA and \userB learn the final path length $Dist$ between them by finding the lowest value of items included in $L$: \vspace{-0.15cm}
$$Dist(\userA,\userB)=Dist(\userB,\userA)=\min_{l_{x}\in L}L \vspace{-0.15cm}$$
If $Dist(\userA,\userB)\leq2$, then  \userA and \userB have common friends between them, thus \distest returns identifiers of all these common friends as in the original \foffinder service. 
While we could use \distest to reveal the first hop identifiers for $Dist>2$, we do not due to the privacy requirements outlined in Section~\ref{subsec:privacy-consideration}.

\section{Analysis}
\label{sec:analysis}

This section presents the analysis of \distest, showing that
it fulfills functional and privacy requirements from Section~\ref{subsec:requirements}.
\newline
\subsection{Correctness}

\medskip
\noindent\textsc{Lemma 1.} 
{\em
If $\ppid \in F^k(\paid)$, then:\vspace{-0.15cm}

\begin{enumerate}
\item There exists a path  $X=\{x_i\}$, for $i \in
\{0,\ldots,k-1\}$, between \userA and \userP, in the social graph. \vspace{-0.15cm}
\item \userA receives the $i^{th}$ order capability $c^{i}_{x_i}$
from every $x_i$ in $X$.\vspace{-0.15cm}
\end{enumerate}
}

\noindent\textsc{Proof.} %
When $\ppid \in F^k(\paid)$, by using the logical implication for the social graph building (see Section~\ref{sec:distance}), it must hold that, in order to include $ID_{x_{i+1}}$ in $F^{i+1}(\paid)$, $ID_{x_i}$ must be included in $F^{i}(\paid)$.
Therefore, we can recursively argue: \vspace{-0.15cm}
{\small
$$ \exists x_{k-1}: ID_{x_{k-1}} \in F^{k-1}(\paid) \land ID_{x_{k-1}} \in F(\ppid)  \vspace{-0.15cm}$$ 
$$ \ldots \vspace{-0.05cm}$$
$$ \exists x_{i+1}: ID_{x_{i+1}} \in F^{i+1}(\paid) \land ID_{x_{i+1}} \in F^{k-i}(\ppid) \vspace{-0.15cm}$$
$$ \exists x_i: ID_{x_i} \in F^{i}(\paid) \land ID_{x_i} \in F^{k-1-i}(\ppid) \vspace{-0.15cm}$$
$$ \ldots \vspace{-0.05cm}$$
$$ \exists x_0: ID_{x_0} \in F(\paid) \land ID_{x_0} \in F^{k-1}(\ppid)$$}\\[-2ex]
\noindent Note that, for every $i \in \{0,\ldots,k-1\}$, there exists a connection to $x_{i-1}$ and $x_{i+1}$, 
thus, $\{x_0,x_1,\ldots,x_{k-1}\}$ form a path $X$ between \userA and \userP. 
Considering $x_i \in X,0 \leq i < k-1$, since $x_i \in F^{i}(\paid)$, then \userA receives $c^{i}_{x_i}$. \qed

\medskip
\noindent\textsc{Theorem 1.} 
{\em
Let there be a path $X=\{x_i\}, i \in \{0,...,d\}, d\geq0$ between
\userA and \userB in the social graph. If path $X$ is discovered
by the \distest discovery protocol, then both \userA and \userB can estimate the exact length $d+2$ of path $X$.
}

\begin{figure}[!t]
\centering
	\begin{subfigure}[c]{0.75\columnwidth}
\centering
		\fbox{\includegraphics[width=1\columnwidth]{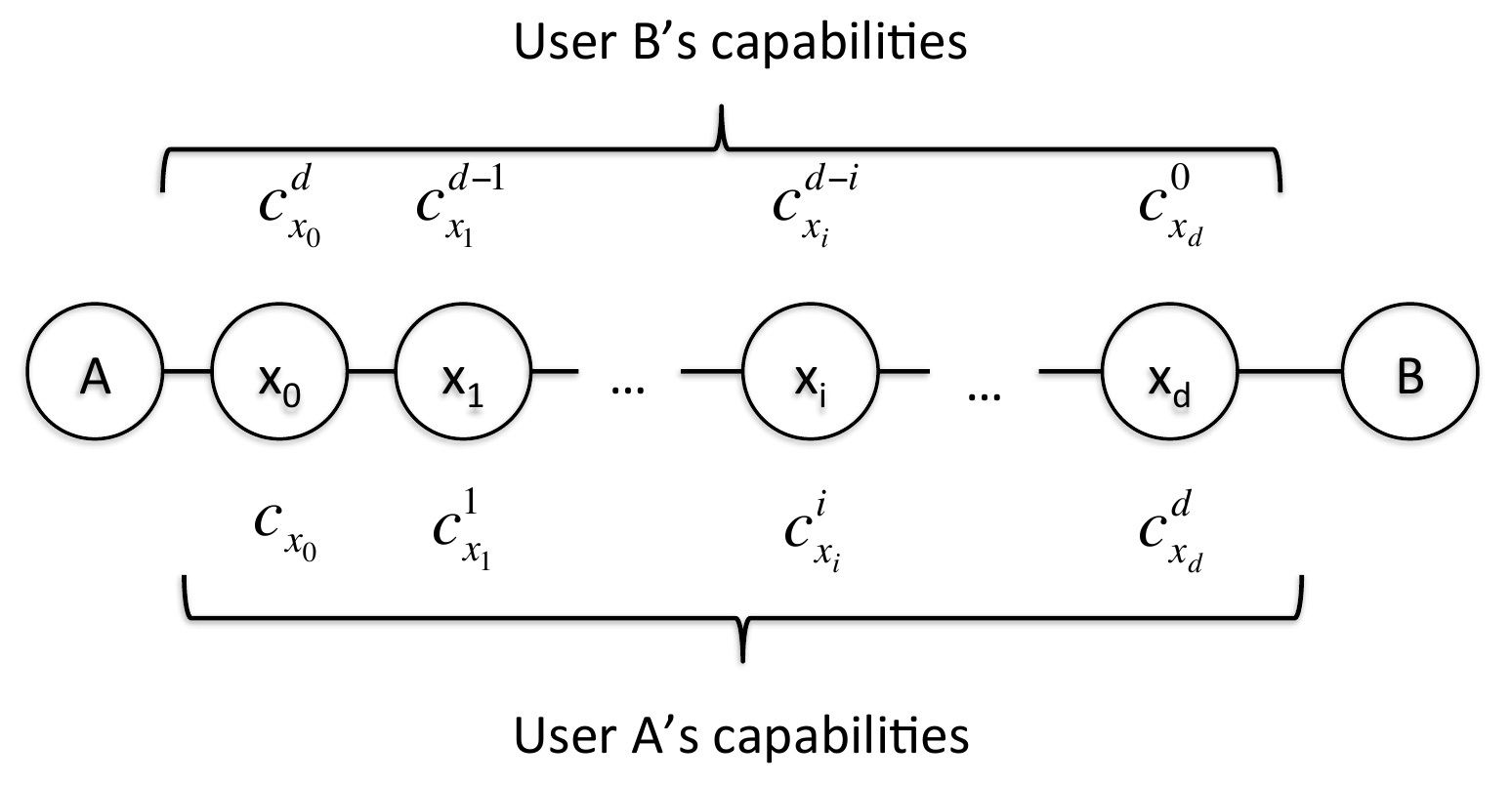}}
		\vspace{-0.1cm}
		\caption{Theorem 1 capability distribution for $d<n$ case.}
                \label{subfig:proof-1}
        \end{subfigure}
\vspace{0.3cm}
\hfill
	\begin{subfigure}[c]{0.75\columnwidth}
\centering
		\fbox{\includegraphics[width=1\columnwidth]{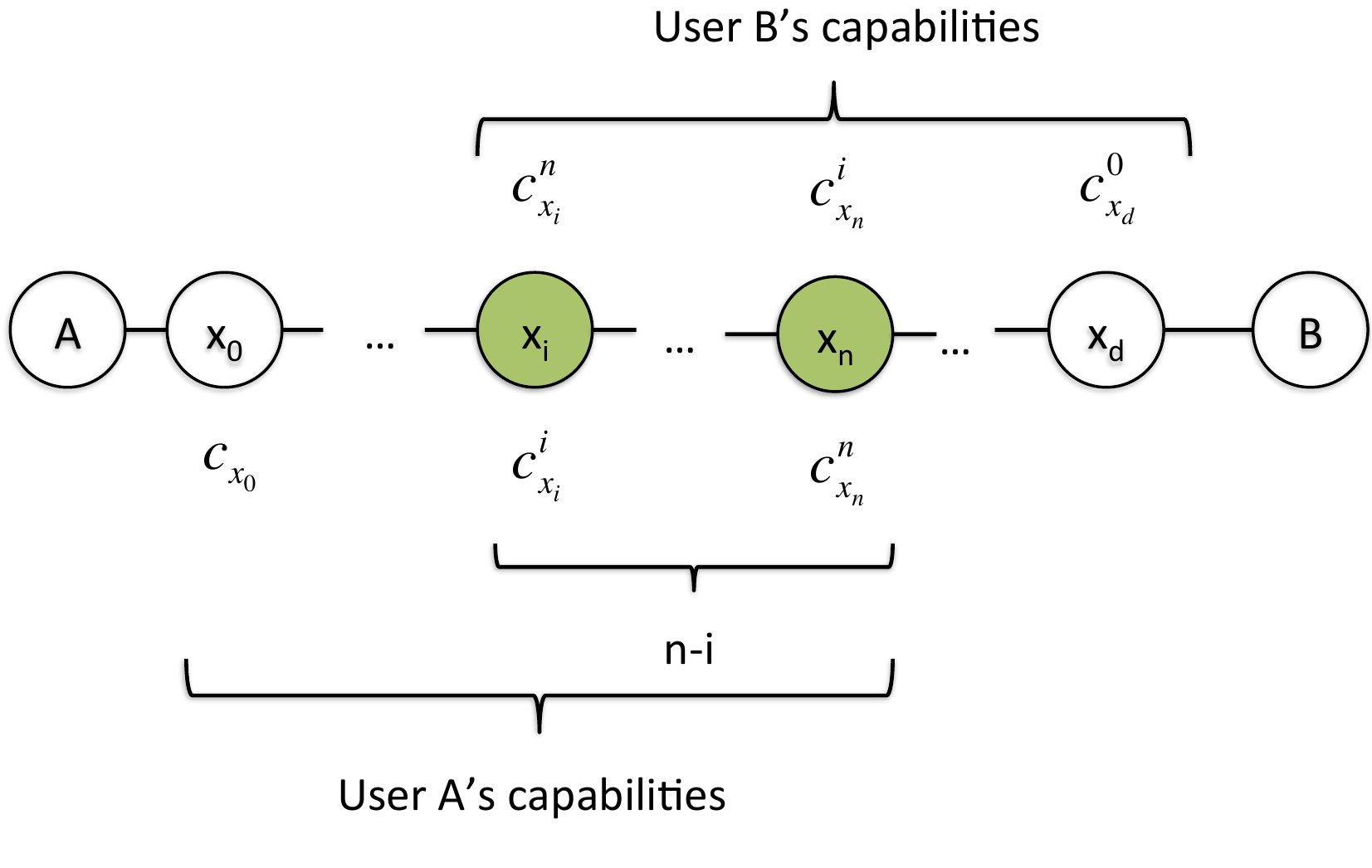}}
		\vspace{-0.1cm}
		\caption{Theorem 1 capability distribution for $d\geq n$ case.}
                \label{subfig:proof-2}
        \end{subfigure}
        		\vspace{-0.1cm}
\caption{Illustration of theorem 1 capability distribution.}
\label{fig:proofs}
\end{figure}

\smallskip
\noindent\textsc{Proof.} %
Let $n$ denote the highest degree of capabilities.\smallskip\\ %
If  $d<n$: \smallskip
\begin{itemize}[leftmargin=0.4cm,noitemsep,nolistsep]
\item[$-$] The set of capabilities of \userA and \userB are $\{c_{x_0},c^{1}_{x_1},\ldots,c^{i}_{x_i},\ldots,c^{d}_{x_d}\}$, and $\{c_{x_d},c^{1}_{x_{d-1}},\ldots,c^{d-i}_{x_{i}},\ldots,c^{d}_{x_0}\}$, respectively.~(See Figure~\ref{subfig:proof-1} for a graphic illustration of the distribution of capabilities for \userA and \userB.)
\item[$-$] If \userA gets a matching capability for $c^{i}_{x_{i}}$, then it must corresponds to $c^{d-i}_{x_{i}}$ for \userB.
\item[$-$] \userA substitutes $k=d-i$ in $Dist(\userA,\userB)$ and receives: \vspace{-0.15cm}
$$l_x=2i+(d-i)+2=i+d+2 \vspace{-0.15cm}$$
\item[$-$] \userA gets multiple capability matches, and sets $Dist(\userA,\userB)$ to be the minimum $l_x$, which is for $i=0$, and $Dist(\userA,\userB)=d+2$. (Similar argument holds for \userB.)
\end{itemize}
\smallskip
If $d \geq n$:\smallskip
\begin{itemize}[leftmargin=0.4cm,noitemsep,nolistsep]
\item[$-$] The set of capabilities of \userA and \userB are $\{c_{x_0},c^{1}_{x_1},\ldots,c^{i}_{x_i},\ldots,c^{n}_{x_n}\}$ 
and $\{c_{x_d},c^{1}_{x_{d-1}},\ldots,c^{d-i}_{x_{n}},\ldots,c^{d-n}_{x_{i}}\}$, respectively -- see Figure~\ref{subfig:proof-2}. (Capabilities for which \userA and \userB obtains matches are marked in green.) 
\item[$-$] If \userA gets a capability match for: $\{c^{i}_{x_{i}},\ldots,c^{n}_{x_n}\}$, 
\userA substitutes $k=n-i$ in $Dist(\userA,\userB)$ and receives: \vspace{-0.15cm}
$$l_x=2i+(n-i)+2=i+n+2\vspace{-0.15cm}$$
\item[$-$] \userA gets multiple capability matches, and sets $Dist(\userA,\userB)$ to be the minimum $l_x$, which is for $i=d-n$, and $Dist(\userA,\userB)=d+2$. (Similar argument holds for \userB.) \qed
\end{itemize}

\subsection{Privacy}\label{subsec:privacy-consideration}

As discussed in Section~\ref{subsec:requirements}, our \distest instantiation
needs to provide users with strong privacy guarantees, i.e., interaction between two users \userA and \userB
does not reveal any information about their non-mutual friends or any other information than they could
discover by gathering information from the standard OSN interface.

\descr{Capability Intersection.} First, we review the security of the common friends discovery protocol from \foffinder~\cite{bfpsi},
since it constitutes the basis of our work.
Its security, in the honest-but-curious model,
reduces to the privacy-preserving computation of set intersection.
That is, privacy stems from the security of the underlying Private Set Intersection (PSI) 
protocol that \foffinder instantiates to privately intersect capabilities and
discover common friends. This is proven by means of indistinguishability between
a real-world execution and an ideal-world execution where a trusted third party receives
the inputs of both parties and outputs the set intersection. %
\cite{bfpsi} uses Bloom Filter based PSI (BFPSI) and, as discussed in Section~\ref{subsec:pcd},
this does not impact security since sets to be intersected are 
random capabilities, thus, high-entropy, non-enumerable items.

\descr{Discovery.} Now observe that the interactive part of the \distest's (social path) discovery protocol --
i.e., the part where information leakage might occur --
mirrors that of \foffinder's discovery protocol.
During the protocol execution, users \userA and \userB engage
in a BFPSI interaction, on input, respectively, $\overline{\Ii}$
and $\overline{\Ir}$, i.e., the sets of their capabilities, and obtain $X$, which
is used to reconstruct the social path between \userA and \userB. %

If \userA and \userB are friends with each other (or have mutual friends), they can
find out the identity of the user(s) corresponding to matching capabilities, thus, 
learning that there exists a social path of length 1 (or 2) {\em and} the identity 
of their mutual friends, but nothing else. In fact, if an adversary could learn the identity
of non-mutual friends, then, we could build an adversary breaking 
the common friends discovery protocol from~\cite{bfpsi} based on BFPSI.
Similarly, if there exists no social path between \userA and \userB, then
the BFPSI interaction does not reveal any information to each other.

On the other hand, if there exists a social path
between \userA and \userB of length $Dist(A,B)>2$, then the matching capabilities are for user nodes
for which \server has removed identifiers \pjid.
Therefore, \userA and \userB do not learn the identity of the users yielding a social
path between them, but only how many.

\descr{Trust in Server \server.} \update{
Each user \user explicitly authorizes \distest to retrieve the set \friendsU of \user's friends. Requesting users to disclose their friend lists is a common practice in social network and smartphone applications. \distest uses this information to have the server \server maintain, distribute, and, in the case of ersatz nodes, create capabilities attesting to the authenticity of friendships. This implies that \server gradually learns the social graph from \distest users,
however, what \server learns is a small subset of what the OSN already knows. Neither \server nor the OSN learn any additional information,
e.g., as opposed to centralized solutions, user locations or interactions between users. }

\descr{Authenticity of capabilities.} In
Section~\ref{subsec:requirements}, we assumed the use of legitimate \distest client applications: as all mobile platforms provide application-private storage, it is reasonable to assume that an adversary on a client device cannot steal the capabilities downloaded on that device by the legitimate client application or otherwise manipulate the input to the protocol. Alternatively, the integrity of the Bloom Filter could be ensured by letting \server sign the Bloom Filter along with the public key of the corresponding \distest client. The BFPSI protocol would then need to be modified accordingly so that each party checks the signature on the other party's Bloom Filter is valid and that the same keypair is used to establish the secure channel.

\section{Coverage evaluation}
\label{sec:evaluation}

We now present an empirical evaluation of
\distest's coverage,
using three publicly available Facebook sample datasets.
Specifically, we analyze how \textit{coverage} attained by the social path discovery 
depends on the fraction of OSN users who join the system,
i.e., the probability that two users discover an existing path between
them in the social network.

\subsection{Datasets Description}
We use three datasets derived from a single dataset,
created by Gjoka et al.~\cite{gjoka10_walkingfb,mgjoka_recommendations_jsac},
using three different sampling techniques:\smallskip

\noindent (1) The \textbf{\sampled}~\cite{social-filter} is our primary dataset.
It contains $500,000$ users, a connected component derived using the ``forest fire" sampling
method~\cite{large-graph-sampling} from the original dataset~\cite{social-filter}. 
As forest fire sampling does not preserve node degree, each
node in this dataset has an average node degree of $30$, which is
significantly less than in the original dataset. To investigate
the effect of the reduced node degree on coverage, we also use the two
more datasets.\smallskip

\noindent (2) The \textbf{\mhrw}~\cite{gjoka10_walkingfb} is built using the \textit{Metropolis-Hastings Random
Walk} (MHRW) %
method with $28$ independent random walks. It contains the friend
lists of $957,359$ users. We call this the set of \textit{sampled
users}. Each of them has an average of $175$ friends, 
including both other sampled users and %
those who were part of the original dataset but that were not sampled -- we call
them {\em outside users}. 
The \mhrw contains a total of $72.2$ million \textit{outside users} (who are
friends of one or more \textit{sampled users}). Because of the
nature of the MHRW sampling, the average number of connections between two \textit{sampled users} in this
set is only 3, thus it is used to evaluate \distest's coverage among poorly connected users.\smallskip
 
\noindent (3) The \textbf{\bfs}~\cite{gjoka10_walkingfb} is built using \textit{Breadth First Search}
(BFS) from $28$ independent BFS traversals. It consists of
$2.2$ million \textit{sampled users}, with an average of $310$
friends. The number of \textit{outside users} %
is $93.8$ million. BFS sampling results in highly connected
subgraphs, and the average number of
connections among \textit{sampled users} is $53$. Thus,
we use the \bfs to measure \distest's coverage among well connected users.

\ifsubmission
\else
\begin{figure}[ttt]
\centering
\includegraphics[width=0.8\columnwidth, trim=0 0.2cm 0cm 0cm]{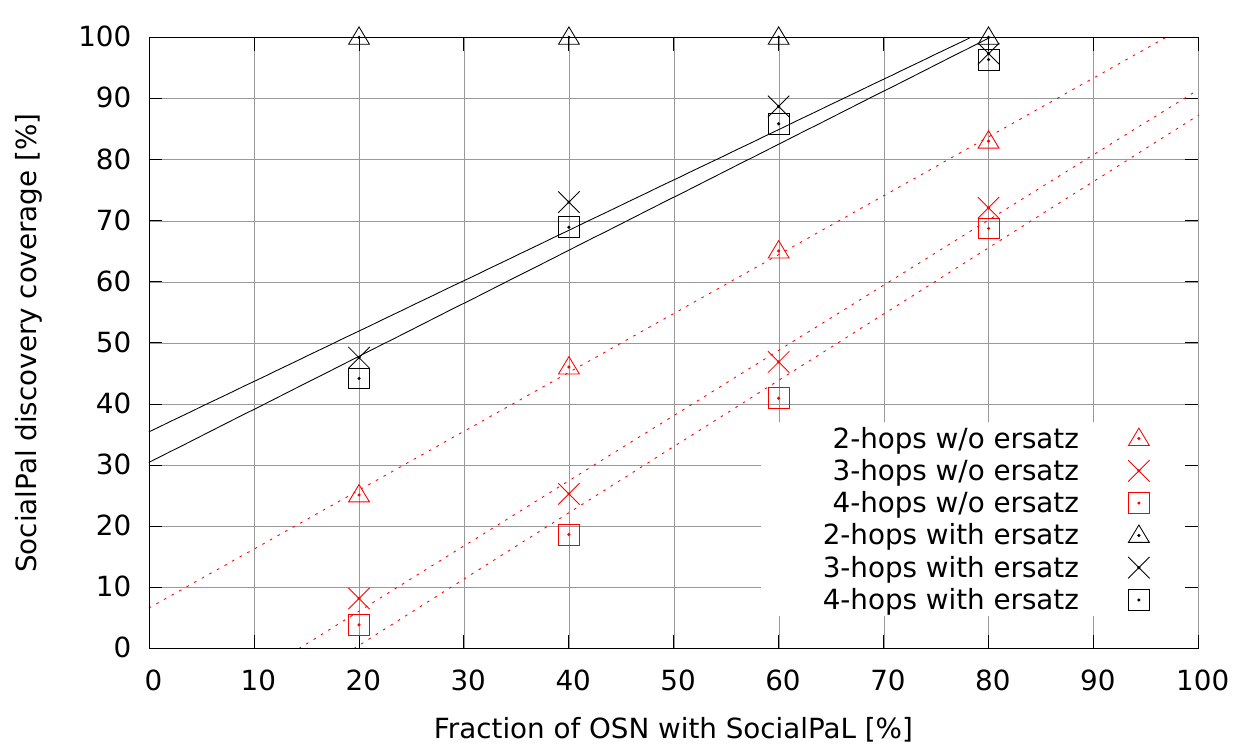}
\caption{Coverage results for the \sampled.} 
\label{fig:user-coverage-1}
\end{figure}
\fi

\subsection{Simulation} %

\noindent{\bf Procedure.} To evaluate \distest's coverage on each of the three datasets, we used the following simulation procedure.
First, we chose, at random, a subset of \textit{sampled users},
which we call the \textit{test set}. For the \sampled, 
we used the whole set as the \textit{sampled users} set.
We chose four different sizes for the {\em test set}: $\{20,40,60,80\}\%$ 
of the \textit{sampled users}. Note that the {\em test set} represents the fraction of the users of an OSN 
who use \distest.

Then, for a given social path length $n$ ($n \in \{2,3,4\}$),
we randomly selected $50,000$ pairs of users from the {\em test set}
in such a way that at least one path of length $n$ exists. 
Finally, we computed the fraction of user pairs for which \distestLine discovers an
existing path between them. We did this for two cases: \distest with
support for ersatz nodes, and without it. Each simulation was repeated 10
times. In total, we conducted 720 different simulations.

\descr{Results.} %
We now present the results of our simulations for each of the three datasets.  
\distest's discovery coverage is presented in 
\ifsubmission \else Figure~\ref{fig:user-coverage-1} and\fi Table~\ref{tab:user-coverage-1} 
for the the \sampled, \ifsubmission \else Figure~\ref{fig:user-coverage-2} and\fi 
Table~\ref{tab:user-coverage-2} for the \mhrw, and 
\ifsubmission \else Figure~\ref{fig:user-coverage-3} and\fi  Table~\ref{tab:user-coverage-3}
for the \bfs. 
\ifsubmission
\else
	Each graph shows how the coverage (for paths of different length) of
	\distest relates to fraction of OSN users who use \distest. Red dotted
	lines indicate the performance of \distest without ersatz node
	support, while black solid lines correspond to the user of ersatz nodes.
\fi

Without ersatz nodes, coverage increases linearly as more users
start using \distest. The rate of growth is highest for the \sampled
and lowest for the \mhrw. In general, the coverage figures are
low. For instance, even if 80\% of OSN users have \distest, the coverage
for paths of length 4 ranges between 0.19\% (the \mhrw) and 68.71\%
(the \sampled).
The introduction of ersatz nodes results in a remarkable improvement 
across the board in all datasets. As expected, the coverage
for paths of length 2 is 100\%. When 80\% of OSN users are in the \distest system,
the coverage is well above 80\% in all cases. Even when only 20\% of
users have \distest, coverage is still above 40\% in all cases, except for the \mhrw. 

\begin{table}[ttt]
	\centering
	\resizebox{0.75\columnwidth}{!}{%
	\begin{tabular}{|c|c|c|c|c|c|c|c|c|c|c|}
	\hline
	\textbf{Fraction of OSN} & \multirow{2}{*}{\textbf{Path length}} & \multicolumn{2}{c}{\textbf{with ersatz [\%]}} & \multicolumn{2}{|c|}{\textbf{w/o ersatz [\%]}} \\ \cline{3-6}
	\textbf{with \distest} & & avg & std & avg & std \\ \hline
	\multirow{3}{*}{\textbf{20\%}} & 2 & 100 & 0.0 & 25.12 & 0.27 \\ \cline{2-6}
	& 3 & 47.59 & 0.11 & 8.17 & 0.09 \\ \cline{2-6}
	& 4 & 44.17 & 0.21 & 3.85 & 0.09 \\ \hline
	\multirow{3}{*}{\textbf{40\%}} & 2 & 100 & 0.0 & 46.05 & 0.30 \\ \cline{2-6}
	& 3 & 73.05 & 0.09 & 25.27 & 0.11 \\ \cline{2-6}
	& 4 & 68.93 & 0.09 & 18.64 & 0.14 \\ \hline
	\multirow{3}{*}{\textbf{60\%}} & 2 & 100 & 0.0 & 65.02 & 0.17 \\ \cline{2-6}
	& 3 & 88.69 & 0.07 & 46.87 & 0.14 \\ \cline{2-6}
	& 4 & 85.86 & 0.11 & 40.94 & 0.19 \\ \hline
	\multirow{3}{*}{\textbf{80\%}} & 2 & 100 & 0.0 & 83.02 & 0.10 \\ \cline{2-6}
	& 3 & 97.32 & 0.04 & 72.10 & 0.18 \\ \cline{2-6}
	& 4 & 96.37 & 0.05 & 68.71 & 0.21 \\ \hline
	\end{tabular}	
	}
	\vspace{-0.15cm}
\ifsubmission
\else      		\caption{Coverage results for the \sampled. (Also see Fig.~\ref{fig:user-coverage-1}).} 
\fi
	\label{tab:user-coverage-1}
\vspace{-0.2cm}
\end{table}
\begin{table}[tt]
	\centering
	\resizebox{0.75\columnwidth}{!}{%
	\begin{tabular}{|c|c|c|c|c|c|c|c|c|c|c|}
	\hline
	\textbf{Fraction of OSN} & \multirow{2}{*}{\textbf{Path length}} & \multicolumn{2}{c}{\textbf{with ersatz [\%]}} & \multicolumn{2}{|c|}{\textbf{w/o ersatz [\%]}} \\ \cline{3-6}
	\textbf{with \distest} & & avg & std & avg & std \\ \hline
	\multirow{3}{*}{\textbf{20\%}} & 2 & 100 & 0.0 & 2.52 & 0.1 \\ \cline{2-6}
	& 3 & 22.77 & 0.23 & 0.15 & 0.02 \\ \cline{2-6}
	& 4 & 27.86 & 0.26 & 0.003 & 0.002 \\ \hline
	\multirow{3}{*}{\textbf{40\%}} & 2 & 100 & 0.0 & 5.26 & 0.15 \\ \cline{2-6}
	& 3 & 43.71 & 0.22 & 0.55 & 0.02 \\ \cline{2-6}
	& 4 & 48.46 & 0.25 & 0.03 & 0.004 \\ \hline
	\multirow{3}{*}{\textbf{60\%}} & 2 & 100 & 0.0 & 7.85 & 0.14 \\ \cline{2-6}
	& 3 & 63.50 & 0.13 & 1.19 & 0.03 \\ \cline{2-6}
	& 4 & 67.09 & 0.24 & 0.09 & 0.005 \\ \hline
	\multirow{3}{*}{\textbf{80\%}} & 2 & 100 & 0.0 &10.34 & 0.14 \\ \cline{2-6}
	& 3 & 82.18 & 0.19 & 2.04 & 0.03 \\ \cline{2-6}
	& 4 & 83.90 & 0.29 & 0.19 & 0.01 \\ \hline
	\end{tabular}	
	}
\vspace{-0.15cm}
\ifsubmission
\else			\caption{Coverage results for the \mhrw. (Also see Fig.~\ref{fig:user-coverage-2}).}
\fi
	\label{tab:user-coverage-2}
\end{table}
\begin{table}[t]
	\centering
	\resizebox{0.75\columnwidth}{!}{%
	\begin{tabular}{|c|c|c|c|c|c|c|c|c|c|c|}
	\hline
	\textbf{Fraction of OSN} & \multirow{2}{*}{\textbf{Path length}} & \multicolumn{2}{c}{\textbf{with ersatz [\%]}} & \multicolumn{2}{|c|}{\textbf{w/o ersatz [\%]}} \\ \cline{3-6}
	\textbf{with \distest} & & avg & std & avg & std \\ \hline
	\multirow{3}{*}{\textbf{20\%}} & 2 & 100 & 0.0 & 13.69 & 0.34 \\ \cline{2-6}
	& 3 & 42.42 & 0.30 & 4.31 & 0.11 \\ \cline{2-6}
	& 4 & 54.11 & 0.21 & 1.51 & 0.04 \\ \hline
	\multirow{3}{*}{\textbf{40\%}} & 2 & 100 & 0.0 & 23.85 & 0.26 \\ \cline{2-6}
	& 3 & 63.46 & 0.23 & 10.98 & 0.17 \\ \cline{2-6}
	& 4 & 72.39 & 0.32 & 6.71 & 0.11 \\ \hline
	\multirow{3}{*}{\textbf{60\%}} & 2 & 100 & 0.0 & 33.19 & 0.26 \\ \cline{2-6}
	& 3 & 78.62 & 0.17 & 18.50 & 0.09 \\ \cline{2-6}
	& 4 & 84.28 & 0.17 & 14.30 & 0.13 \\ \hline
	\multirow{3}{*}{\textbf{80\%}} & 2 & 100 & 0.0 & 41.46 & 0.28 \\ \cline{2-6}
	& 3 & 90.39 & 0.12 & 26.24 & 0.19 \\ \cline{2-6}
	& 4 & 93.03 & 0.10 & 22.71 & 0.11 \\ \hline
	\end{tabular}	
	}
	\vspace{-0.15cm}
\ifsubmission
\else			\caption{Coverage results for the \bfs. (Also see Fig.~\ref{fig:user-coverage-3}).} 
\fi
	\label{tab:user-coverage-3}
\end{table}

\subsection{Take-aways}

\descr{Ersatz nodes dramatically improve coverage.}
With ersatz nodes, \distest
discovers 100\% of social paths of length 2,
thus addressing one of the major limitations of the \foffinder system~\cite{bfpsi}.
The coverage for paths of length 3 and 4 always increases, between
$10\%$ and $80\%$, depending on the fraction of OSN users in
\distest and the dataset used for the simulations.

\ifsubmission
\else
\begin{figure}[!t]
\centering
\includegraphics[width=0.8\columnwidth, trim=0 0.2cm 0cm 0cm]{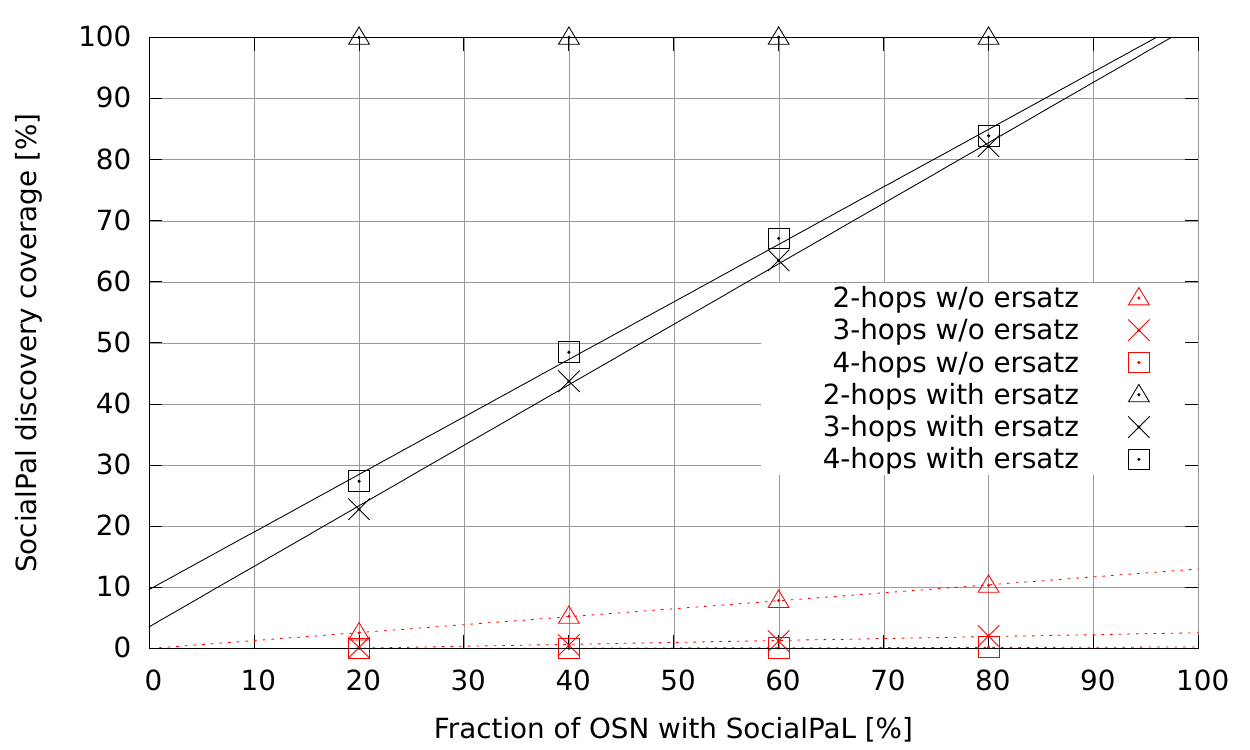}
\caption{Coverage results for the \mhrw.} %
\label{fig:user-coverage-2}
\end{figure}
\fi

\descr{Variation of coverage across different datasets.} We observe
better coverage results with the \bfs than with the \mhrw. As the \bfs
represents coverage among well connected users, the density of
ersatz nodes between random users is higher than in the \mhrw,
thus yielding better overall coverage. The \bfs models
societies, such as most of the western societies, where the
penetration of OSNs is high. The high coverage results with
the \bfs suggests that \distest will do well in this context. On
the other hand, the \mhrw models societies where OSN connectivity
is poor and, although \distest is not as effective here, 
it may still perform reasonably well, detecting
the majority of social paths even before the number of users
joining \distest reaches 50\%.

\ifsubmission
\else
\begin{figure}[t]
\centering
\includegraphics[width=0.8\columnwidth, trim=0 0.2cm 0cm 0cm]{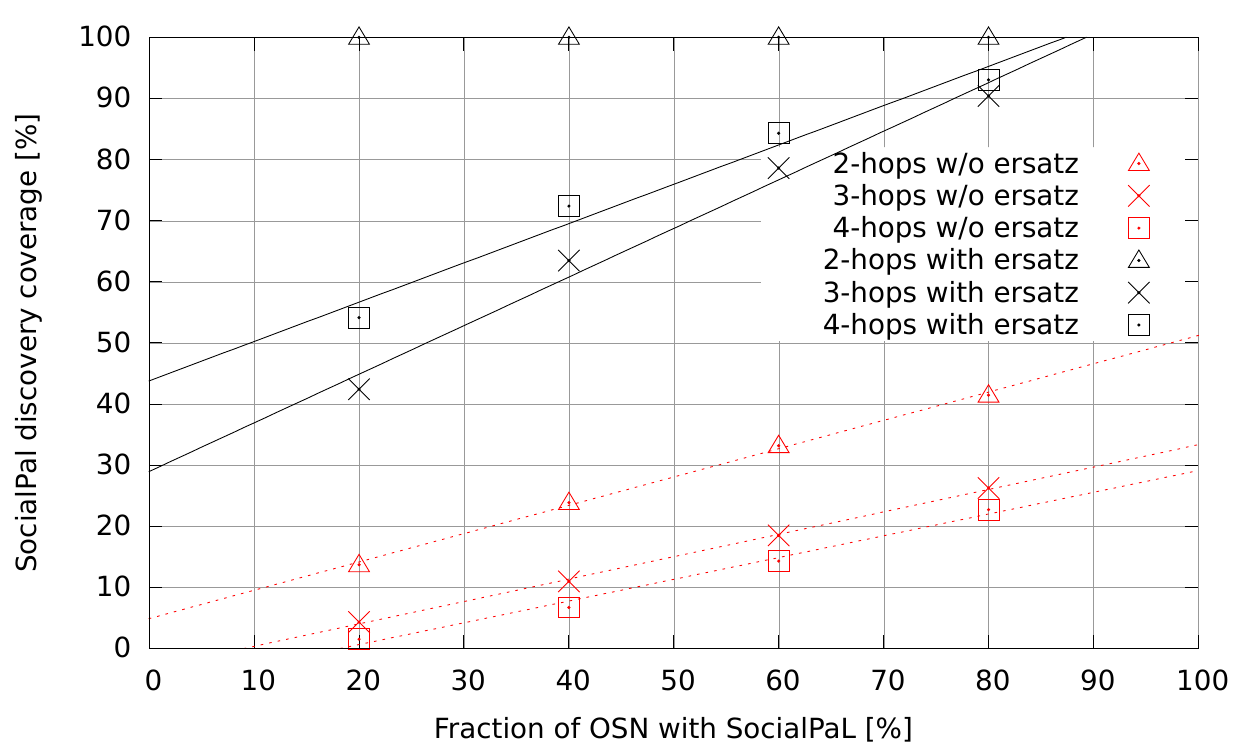}
\caption{Coverage results for the \bfs.} %
\label{fig:user-coverage-3}
\vspace{-0.1cm}
\end{figure}
\fi

\section{Implementing Social Pal}
\label{sec:implementation}

In this section, we present our full-blown implementation of the \distest system.
We aim to support scalability for increasing number of users (in terms of CPU performance and memory)
and to enable developers to easily integrate it
into their applications.

\subsection{Server Architecture}
\label{sec:server-arch}

\descr{Server components.} On the server side, the \distest system extends the \peershare server~\peersharerefonly, which allows two or more users to share sensitive data among social contacts, e.g., friends in a social network. %
We use the following basic functions of \peershare: (1) OSN interfaces to retrieve  social graph information, (2) the OAuth~\cite{hammer2011oauth} component for user authentication, and (3) the data distribution mechanism. On top of these components, we develop a new server architecture that supports the addition of server-based applications via an extension mechanism. This design choice allows us to implement the \distest functionality in such a way that the system can efficiently scale (in terms of memory and CPU performance) to support an increasingly large number of users.

\begin{figure}[!t]
\centering
\includegraphics[width=0.85\columnwidth, trim=0 0cm 0cm 0cm]{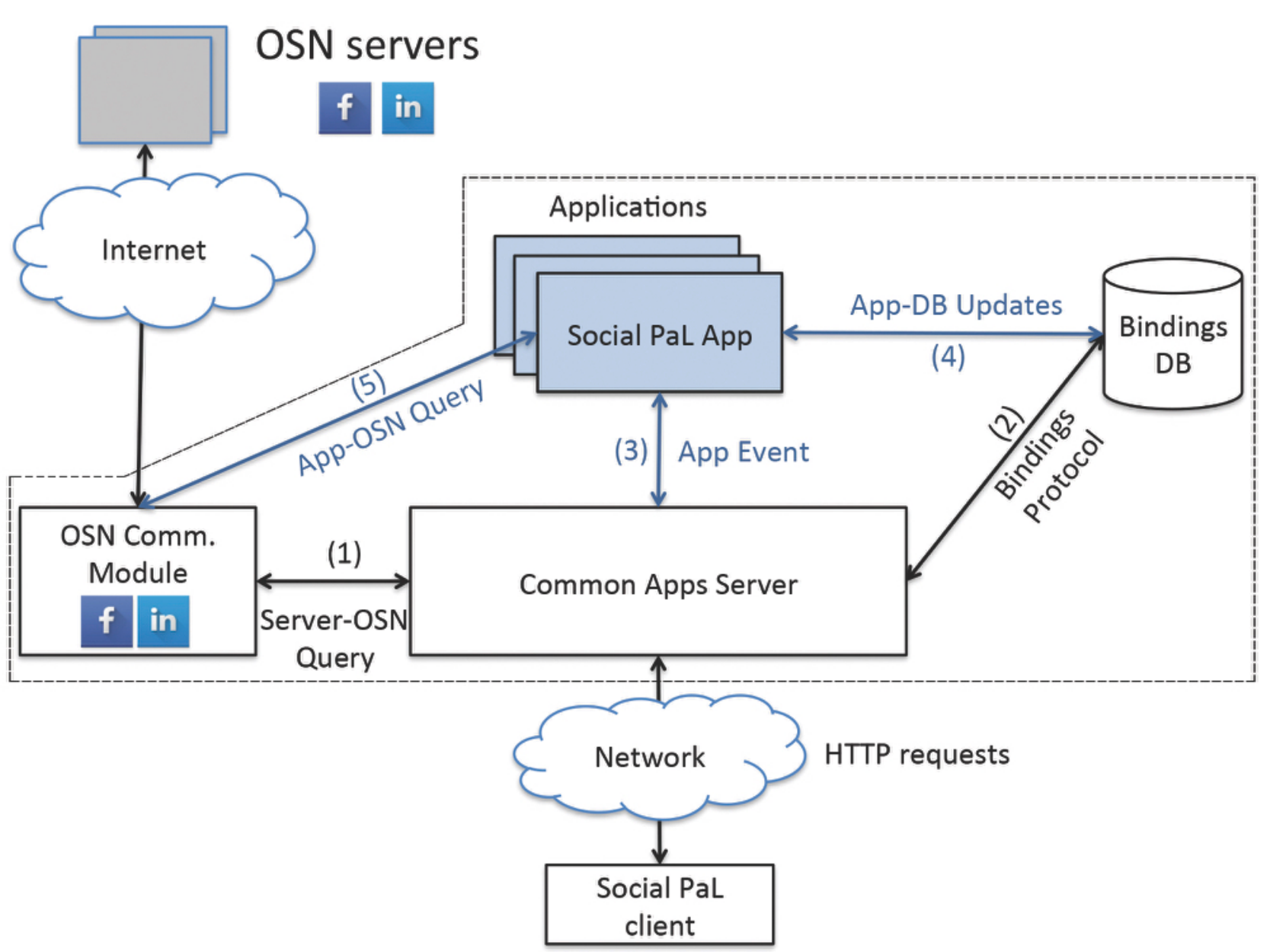}
\vspace{-0.2cm}
\caption{Single instance of \distest server architecture. Components common to all applications are presented in black, while application specific elements are presented in blue.}
\vspace{-0.2cm}
\label{fig:server-arch}
\end{figure}

As illustrated in Figure~\ref{fig:server-arch}, the server architecture consists of the following components: the \textit{\basicserver}, a group of applications (e.g., the \distest App), the \textit{OSN Communicator Module}, and the \textit{Bindings Database} (Bindings DB). The \basicserver provides the basic functionality that is common for all applications: (1) storage of data uploaded by users in the Bindings DB, (2) distribution of users' data to other authorized users, and (3) retrieval of basic social graph information, %
which is needed for enforcing the appropriate data distribution policy. The OSN Communicator Module is a plugin-based service responsible for querying OSNs for social graph information. Its plugin-based structure allows us to easily add support for new OSNs. The Bindings DB stores data uploaded by users and information on how to distribute them among social contacts. 

\ifsubmission
\else
The components interact with each other using a number of different interfaces: \textit{Server-OSN Query}, \textit{Bindings Protocol}, \textit{\appiface}, \textit{\dbupdates} and \textit{App-OSN Query} (cf. Fig~\ref{fig:server-arch}). The Server-OSN Query interface is used by the \basicserver to retrieve social graph information from the OSN. The Bindings Protocol interface specifies communication between the \basicserver and the Bindings DB. These two interfaces provide basic data distribution functionality to all applications. The other three interfaces are used only by applications that need to perform specific modifications on distributed data on the server side. 

Each application that requires logic on the server side has to be registered in the \basicserver via a Uniform Resource Identifier (URI). This URI is used by the \appiface callback interface to notify the application about incoming application-specific events (e.g., an upload of new data items). In addition, this interface may be used to modify data read from the Bindings DB before returning them via the \basicserver to the user (e.g., generation of higher order capabilities). The server application itself has access to the two remaining interfaces. It uses the \dbupdates to modify data stored in the Bindings DB or to create new data items. The App-OSN Query interface is used when the application needs to obtain social graph information from the OSN.
\fi

\begin{figure}[!t]
\centering
\includegraphics[width=0.85\columnwidth, trim=0cm 0cm 0cm 0cm]{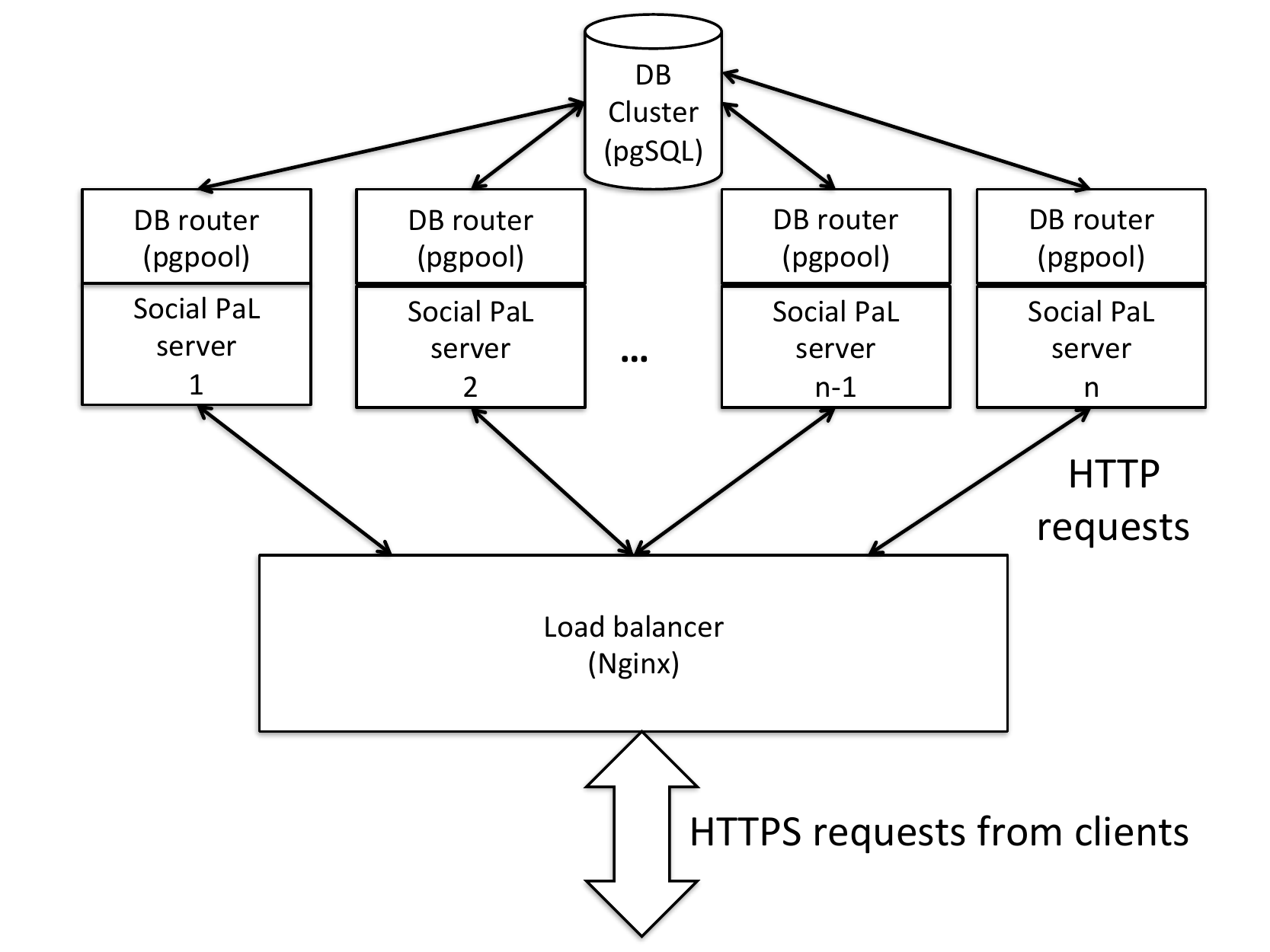}
\vspace{-0.25cm}
\caption{Scalable \distest server architecture.}
\label{fig:server-scaling}
\vspace{-0.15cm}
\end{figure}

\descr{\distest server implementation.} The \distest server application is notified via 
\ifsubmission\else the \fi \appiface  callback interface about new capabilities uploaded by users. It uses 
\ifsubmission\else the \fi \dbupdates interface to create any required ersatz nodes and properly update the recipient sets of capabilities during the social graph building process. The \distest application also uses the \appiface interface to handle capability download requests. The \basicserver, instead of immediately returning data it has read from the Bindings DB, passes them to the \distest application that generates any missing higher order capabilities. Finally, the \distest application returns the complete set of capabilities back to the \basicserver, which completes the request handling.
Note that our implementation of the OSN Communicator Module supports interactions with both LinkedIn and Facebook.

\descr{Implementation details.}
Our core \distest server is written in PHP. We support capabilities of $0^{th}$ and $1^{st}$ order, allowing to discover social paths between users that are up to $4$ hops from one another. Based on relevant prior work~\cite{backstrom2011,milgram1967small,facebook-anatomy}, a $4$-hop distance is enough for most practical use cases. As the Bindings DB needs to store the capabilities of users and information about how to share those, the necessary amount of persistent storage will substantially grow if \distest becomes widely used. Thus, to limit the data storage overhead, \distest server does not store any higher order capabilities in the Bindings DB, but generates them when requested by the requesting client. Tests on our server show that generating higher order capabilities has a negligible impact on the \distest \textit{capability distribution protocol} performance (i.e., the server generates 1 million higher order capabilities in about 500ms using the hardware described in Section~\ref{sec:server-eval}). 
Finally, in order to implement the LinkedIn OAuth module for the OSN Communicator Module, we use the OAuth Pecl extension for PHP, %
while, for the Bindings DB, PostgreSQL database server. %

\descr{System scaling.}
Since \distest may generate a large number of server requests if used by a large number of users, we can take following steps to ensure that the system can scale. Our proposed scaling architecture is illustrated in Figure~\ref{fig:server-scaling}. It includes a powerful HTTP front-end server (such as  Nginx) %
acting as load balancer, which terminates incoming secure HTTPS connections and forwards server requests upstream to $n$ instances of \distest servers acting as request handlers. Each \distest server instance will run the \textit{HipHop Virtual Machine} (HHVM) %
daemon that handles HTTP requests. HHVM usage can massively improve server performance, as it uses just-in-time compilation to take advantage of the native code execution instead of the interpreted one~\cite{hhvm-php}. 

Each instance of \distest server runs, locally, a database query router (pgpool) %
providing access to the actual database cluster including multiple PostgreSQL servers. The query router enhances the overall database access performance by keeping open connections to the database cluster, load-balancing the stored data among multiple instances of the database servers, and temporarily queuing requests for database access in case of cluster overload. Note that there are no cross-dependencies between the \distest server instances for the database read access, thus, no complex control mechanism is needed to support this parallelism.

\descr{Server code.}  The source code of the server implementation 
is available from \url{https://github.com/SecureS-Aalto/SoPaL}.

\vspace{0.22cm}
\subsection{Server Performance Evaluation}\label{sec:server-eval}

\descr{Performance testbed.}
We evaluated our server implementation in a testbed consisting of two machines: the first played the role of a single \distest server instance (cf. Figure~\ref{fig:server-arch}), while the second simulated a group of client devices. The server ran on a 4-core machine with a 2.93GHz CPU on each core and 128GB of RAM. It hosted Nginx (version 1.1.19), PostgreSQL server (version 9.1), and php5-fpm %
for the PHP 5.6 engine. 
\ifsubmission
\else
To improve the overall server performance, we adjusted the default settings for Nginx, PostgreSQL and php5-fpm (see Table~\ref{tab:server-settings}).

\begin{table}[t]
\centering
\scriptsize
\begin{tabular}{|c|c|}
\hline
\bf{System parameter} & \bf{Value} \\ \hline
\multicolumn{2}{|c|}{\cellcolor{lightgray}\emph{Nginx}} \\ \hline
worker\_processes & $4$ \\ \hline
worker\_connections & $2048$ \\ \hline
client\_body\_timeout & $12$ \\ \hline
client\_header\_timeout & $12$ \\ \hline
send\_timeout & $10$ \\ \hline
\multicolumn{2}{|c|}{\cellcolor{lightgray}\emph{PostgreSQL}} \\ \hline
max\_connections & $400$ \\ \hline
shared\_buffers & $4$GB \\ \hline
temp\_buffers & $16$MB \\ \hline
work\_mem & $5$MB \\ \hline
seq\_scan & off \\ \hline
\multicolumn{2}{|c|}{\cellcolor{lightgray}\emph{Network stack}} \\ \hline
net.ipv4.tcp\_window\_scaling & $1$ \\ \hline
net.core.rmem\_max & $16777216$ \\ \hline
net.core.wmem\_max & $16777216$ \\ \hline
net.ipv4.tcp\_rmem & $4096$ $87380$ $16777216$ \\ \hline
net.ipv4.tcp\_wmem & $4096$ $16384$ $16777216$ \\ \hline
\end{tabular}
\vspace{-0.15cm}
  \caption{Details of server implementation settings.}
  \label{tab:server-settings}
\ifsubmission\vspace{-0.4cm}\fi
\end{table}
 
\fi
Inter-process communication was implemented via UNIX  sockets. The machine running the clients had 8 CPU cores (at 2.93 GHz) and 64GB of RAM.

To eliminate the unpredictability of network latency, we modified the server implementation by replacing the OSN Communicator Module with the local service that provided social graph information based on the \mhrw. %
We populated the Bindings DB with capabilities generated for the 120,000 \textit{sampled users} from the \mhrw. The capabilities of \textit{sampled users} together with capabilities generated for ersatz nodes constituted about 10 million data items that were stored in the Bindings DB. Finally, to minimize impact of the client-server transmission delay, we kept the server and client machines in the same network and connected them using a 1 Gbit/s Ethernet link via the single switch.

\descr{Experiments.} 
We evaluated server performance by sending bursts of $n$ requests per second, for $n\in \{1,10,20,30,40,50,60,70\}$, from the client machine to fetch capabilities from the server (i.e., download of \Ru and \Rud in Figure~\ref{fig:distance-capability-protocol}) for $60$ seconds. Fetching capabilities involves many read operations on the Bindings DB, thus yielding the highest load on the server among all operations of the \distest capability distribution protocol. %
In each experiment, which we repeated ten times, we measured the number of received responses per second together with the latency of each response on the client machine, and CPU usage together with memory consumption on the server machine. 

\begin{figure}[!t]
\centering
	\begin{subfigure}[b]{\columnwidth}
	\centering
		\includegraphics[width=0.78\textwidth,trim=1.5cm 0 1.5cm 0]{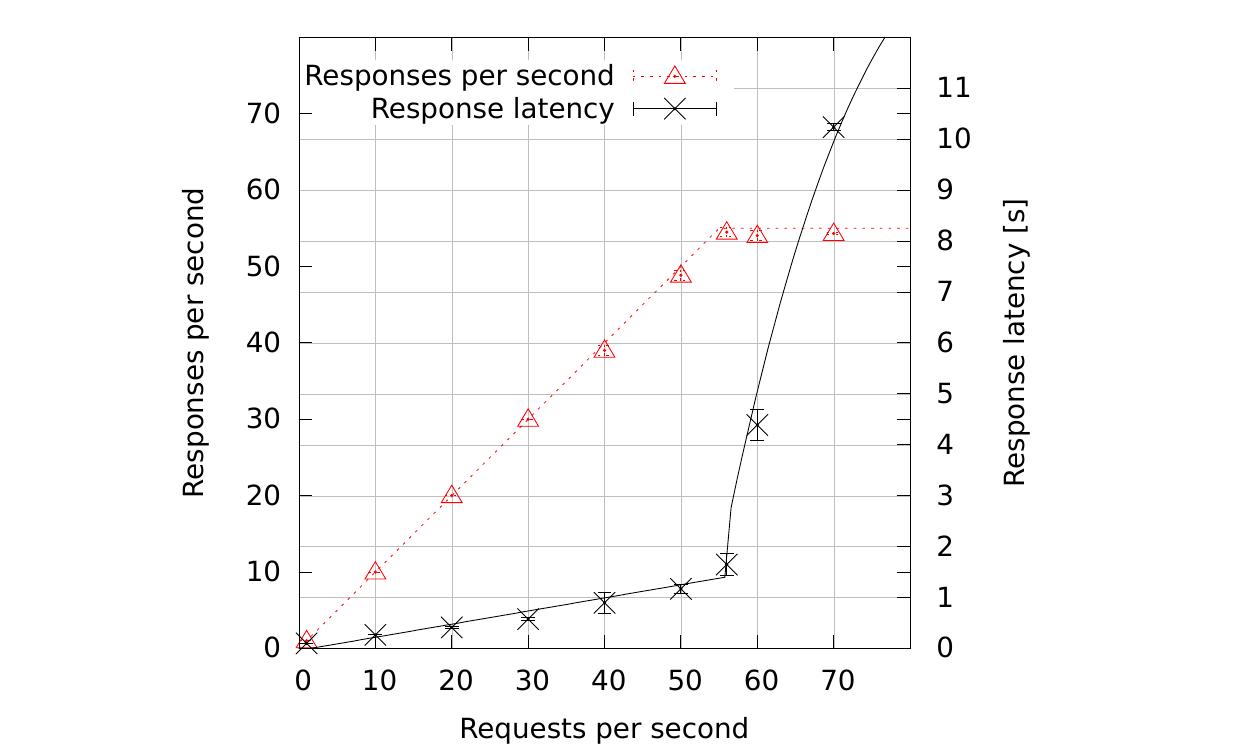}
		\caption{Responses per second and response latency vs \# requests per second.}
                \label{subfig:recv-latency}
	\end{subfigure}
        \vspace{-0.05cm}
      	\begin{subfigure}[b]{\columnwidth}
	\centering		
		\includegraphics[width=0.78\textwidth]{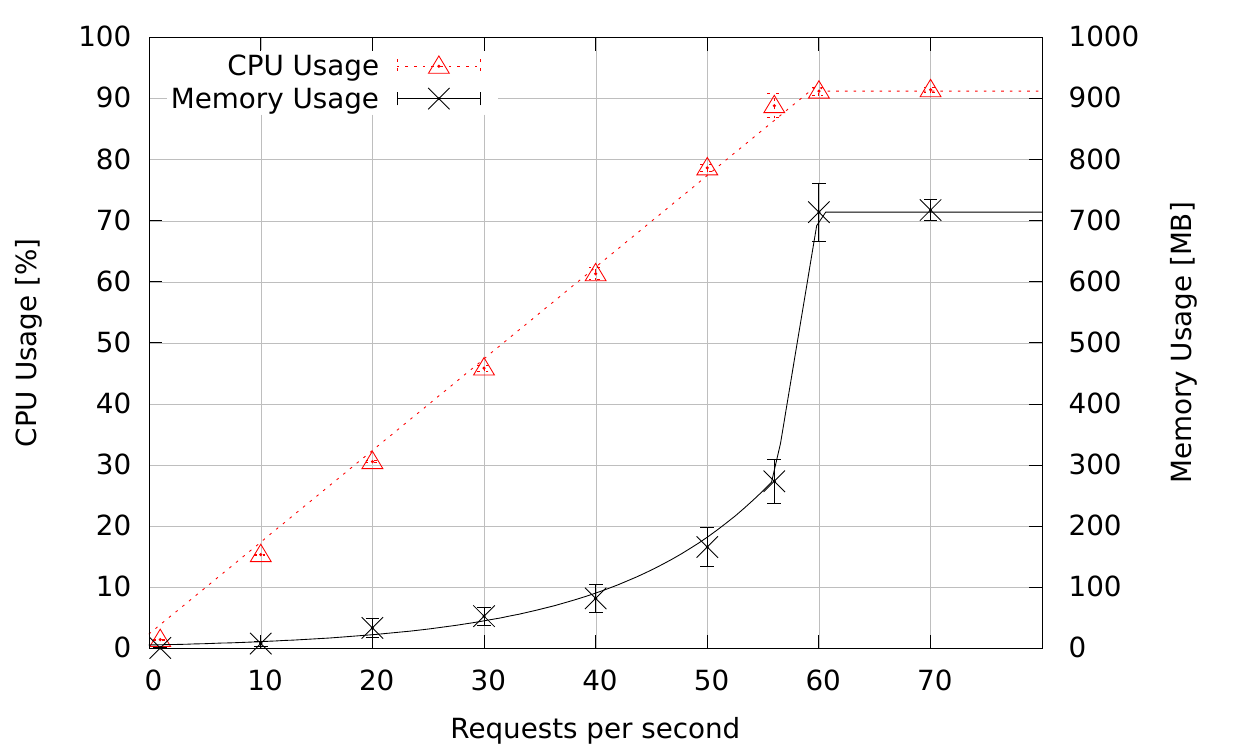}
		\caption{CPU usage and memory consumption vs \# requests per second.}
                \label{subfig:cpu-ram}
        \end{subfigure}
\vspace{-0.5cm}
\caption{\distest server performance.}
\label{fig:server-eval}
\end{figure}

\descr{Results.} Figure~\ref{fig:server-eval} illustrates the results of our experiments. We observe that $56$ requests per second yields a saturation point for the server. Below $56$ requests/second, the number of responses per second and the response latency grow linearly. Whereas, as depicted in Figure~\ref{subfig:recv-latency}, above $56$,  we observe an exponential growth of the response latency and the constant number of received responses per second.
Figure~\ref{subfig:cpu-ram} also shows that CPU usage reaches more than $90\%$ above $56$ requests/second. Peak memory consumption is about $700$MB, which also shoots up significantly when the number of requests crosses the saturation point. %

The $56$ requests/second saturation point shows that the performance of our server implementation is in line with that emerging from studies of systems equipped with similar hardware~\cite{hhvm-php,hhvm-php-2}. %
We also looked at the server performance when only handling $1$ request per second and observed that the average response latency is about $50$ms, and the average Bindings DB interaction time is around $5$ms. Since client-server network latency is negligible, the vast majority of request handling takes place in the PHP interpreter,
which highlights that PHP is the server's bottleneck. 
Therefore, in order to improve the server performance, php5-fpm should be replaced with HHVM, which is reported to be significantly more performant~\cite{hhvm-php,hhvm-php-2}. Further gains could also be obtained by migrating the PostgreSQL server to a separate machine connected over a fast link (cf. Fig.~\ref{fig:server-scaling}). We leave these as part of future work.

Assuming that the server handles $56$ requests per second, a total of $4.84$ million requests can be processed daily by a single instance of \distest server with comparable hardware capabilities. Assuming that each user executes the \distest capability distribution protocol around $4$ times a day, about $1.21$ million \distest users can be handled by one \distest server instance. Since user requests are independent of each other, and because the scalable architecture of the \distest server allows adding further instances easily (as described in Section~\ref{sec:server-arch}), the total capacity of \distest system amounts to the cumulative number of users that can be handled across all \distest server instances. Finally in order to avoid making PostgreSQL become the bottleneck of the system (which may be caused if many \distest server instances are added), the Bindings DB should be turned into a database cluster with data sharding and replication enabled. This guarantees that the data kept in the Bindings DB is synchronized and accessible with high enough availability.

\vspace{0.2cm}
\subsection{Client Implementation}
\label{sec:framework}

\begin{figure}[!t]
\centering
\includegraphics[width=0.91\columnwidth, trim=0 0cm 0cm 0cm]{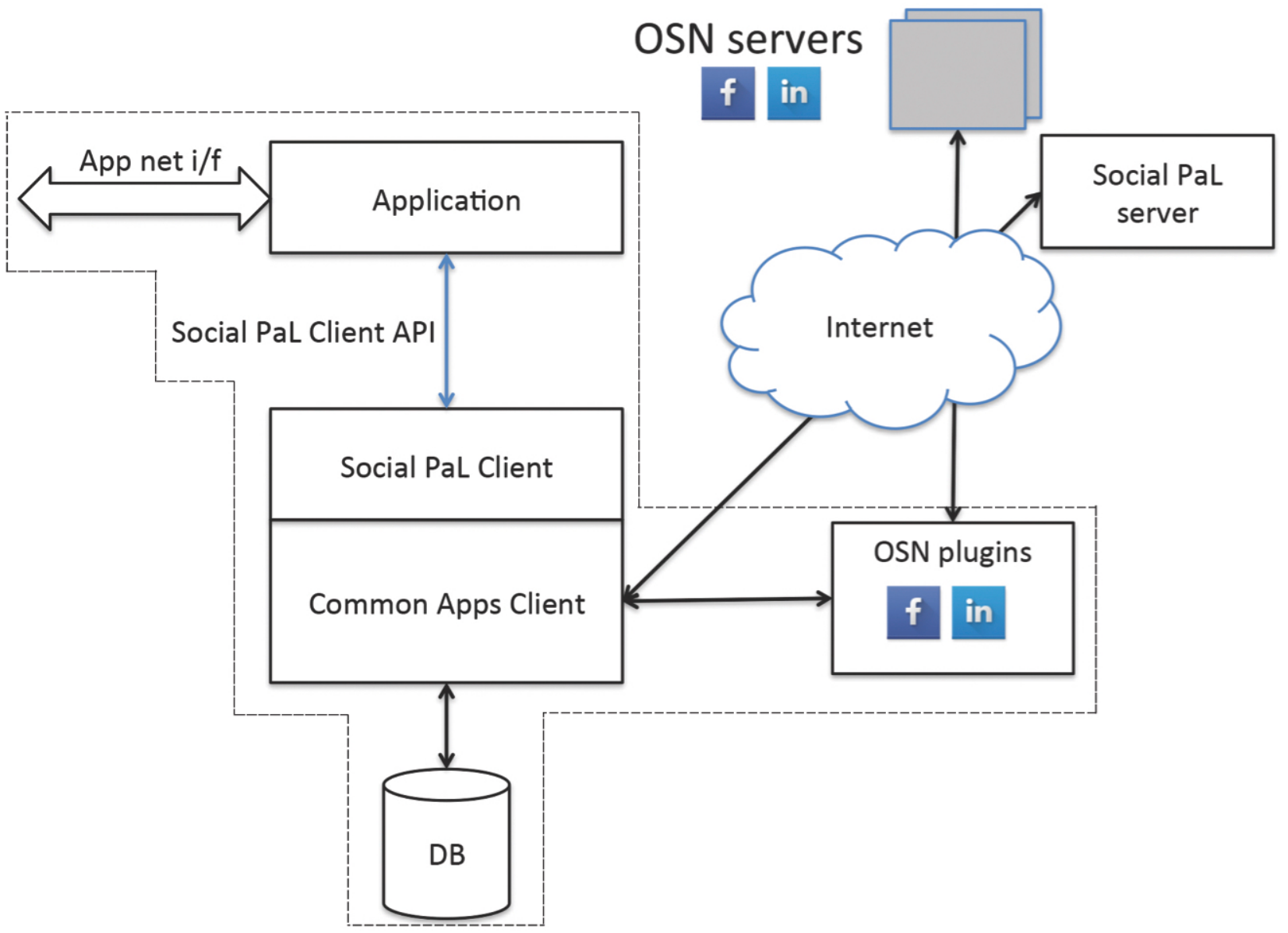}
\vspace{-0.15cm}
\caption{\distest client architecture.}
\label{fig:client-arch}
\end{figure}

\descr{Client architecture.} The client-side architecture of \distest is depicted in Figure~\ref{fig:client-arch}. It consists of the {\em Common Apps Client}, the set of OSN plugins, and the {\em \distest Client}. The first two components are responsible for the communication with the \distest server, while the last one provides the interface for the applications. Together, these components form a mobile platform library that can be easily imported by developers into their applications. To facilitate support for multiple OSNs, similar to our server design, we have decoupled OSN-specific functionality from the Common Apps Client and made it a plugin-based solution. 

We have considered two possible design choices for the client architecture: (1) designing it as a stand-alone service with applications connecting to it, using available inter-process communication mechanisms, or (2) as a service integrated into the application. We choose the latter as it supports application-private storage for capabilities (i.e., not accessible by other applications) and enables each application to have its own \distest server. This choice provides additional protection against capability leakage to a malicious application and removes the requirement to deploy the global \distest server. On the other hand, if multiple instances of \distest application runs on the same device, 
we would incur increased network traffic and require more storage space in comparison to the stand-alone service approach. 
We argue that this tradeoff is acceptable, as the \distest capability distribution is run no more than a few times a day. Also, the amount of data to store is likely to be limited in the order of tens of megabytes, which is justifiable given the clear usability and deployability advantages.

\descr{Implementation details and performance.} We have implemented the client library on Android, operating as an Android lightweight service. 
\ifsubmission
\else 
To evaluate the performance on the client, we measured running times of the \distest discovery protocolon a Samsung Galaxy Tablet GT-P3100 running Android 4.1.2 API 16 and a ZTE Blade S6 running Android 5.0 API 21 connected via Bluetooth.
We assumed that both parties have the same number of input items, ranging from 1000 to 35000 (with 5000 increments). We also fixed the intersection of the sets to be $2.5\%$ of the set size. Table~\ref{tab:perf} shows average computation and communication times.

\begin{table}[t!]
	\centering
	\resizebox{0.675\columnwidth}{!}{%
	\begin{tabular}{|c|c|c|c|c|}
	\hline
	\multirow{2}{*}{\textbf{Input size}}  & \multicolumn{2}{c}{\textbf{Computation [s]}} & \multicolumn{2}{|c|}{\textbf{Communication [s]}} \\ \cline{2-5}
	& avg & std & avg & std \\ \hline
	1000 & 2.24 & 0.09 & 1.51 & 0.02 \\ \hline
	5000 & 10.15 & 0.21 & 3.8 & 0.11 \\ \hline
	10000 & 19.55 & 0.48 & 7.17 & 0.28 \\ \hline
	15000 & 29.22 & 0.55 & 11.84 & 0.35 \\ \hline
	20000 & 39.79 & 0.95 & 13.82 & 0.38 \\ \hline
	25000 & 52.2 & 0.65 & 17.06 & 0.17 \\ \hline
	30000 & 63.44 & 1.3 & 20.38 & 0.16 \\ \hline
	35000 & 77.0 & 1.93 & 23.86 & 0.06 \\ \hline
	\end{tabular}	
	}
	\vspace{-0.15cm}
        	\caption{Computation and communication time (in seconds) for the \distest discovery protocol for increasing input set sizes.        \label{tab:perf}}
\end{table}

\fi

\descr{\distest Client API.} The \distest application interface is used by applications to run the \distest discovery protocol.
It has been designed to be readily usable by application developers that are not cryptography experts, but are nonetheless interested in implementing privacy-preserving discovery of social paths. This allows developers to delegate the responsibility of this process to the \distest Client, and requires them to integrate only four basic methods into the application code, which we present below. 

Applications running the \distest discovery protocol act as \distest message relays between the two \distest Client instances. The application starting the \distest discovery session calls the \texttt{startSoPaLSession} method. This returns an opaque \distest object, which is forwarded to the remote party. From this point onward, both parties invoke \texttt{handleSoPaLMessage} for every message received. This method processes the received message, and if needed, creates a response. It also returns a flag indicating if the protocol execution is completed. If so, the application uses the \texttt{getResult} method to get the social path length it has to a remote party. Finally, the application must call \texttt{endSoPaLSession} to let the \distest Client release all resources from the session.

Besides these four basic methods, the \distest Client also provides three advanced methods: (1) \texttt{rejectSoPaLSession} creates a \distest message that can be sent by the application to the remote party if it does not want to run the discovery protocol; (2) \texttt{updateCapabilities} and (3) \texttt{renewCapability} can be used by the application to force fetching the most recent capabilities from the server, and to generate and upload a new capability to the server, respectively.
\ifsubmission
\else
Table~\ref{tab:sopal-iface} summarizes all the methods available in \distest Client.

\begin{table*}[!ht]
\centering
\resizebox{0.78\textwidth}{!}{%
\begin{tabular}{|l|c|c|l|}
\hline
\bf{Name} & \bf{Input} & \bf{Output} & \bf{Description} \\ \hline
\multicolumn{4}{|c|}{\cellcolor{lightgray}\emph{Basic methods}} \\ \hline
\texttt{startSoPaLSession} & {deviceID} & {SoPaL object} & Initiates \distest discovery session \\ \hline
\texttt{handleSoPaLMessage} & {SoPaL object} & {(SoPaL object,true/false)} & Processes received message and creates response to it \\ \hline
\texttt{getResult} & {deviceID} & {value} & Returns final result of \distest discovery protocol \\ \hline
\texttt{endSoPaLSession} & {deviceID} & {true/false} & Releases resources with given \distest discovery session \\ \hline
\multicolumn{4}{|c|}{\cellcolor{lightgray}\emph{Advanced methods}} \\ \hline
\texttt{rejectSoPaLSession} & - & {object} & Returns a generic message notifying the remote party \\
& & & that it doesn't want to run \distest \\ \hline
\texttt{updateCapabilities} & - & Asynchronous & Fetches most recent capabilities from the server \\ \hline
\texttt{renewCapability} & - & Asynchronous & Generates a new capability and uploads it  to the server \\ \hline
\end{tabular}
}
\vspace{-0.15cm}
  \caption{Summary of methods available in the \distest Client.}
  \label{tab:sopal-iface}
\end{table*}
\fi

\section{Sample Applications}
\label{sec:apps}

\ifsubmission
\else
\begin{figure*}[!t]
\centering
	\begin{subfigure}[b]{\columnwidth}
		\centering 	
		\includegraphics[width=0.9\textwidth]{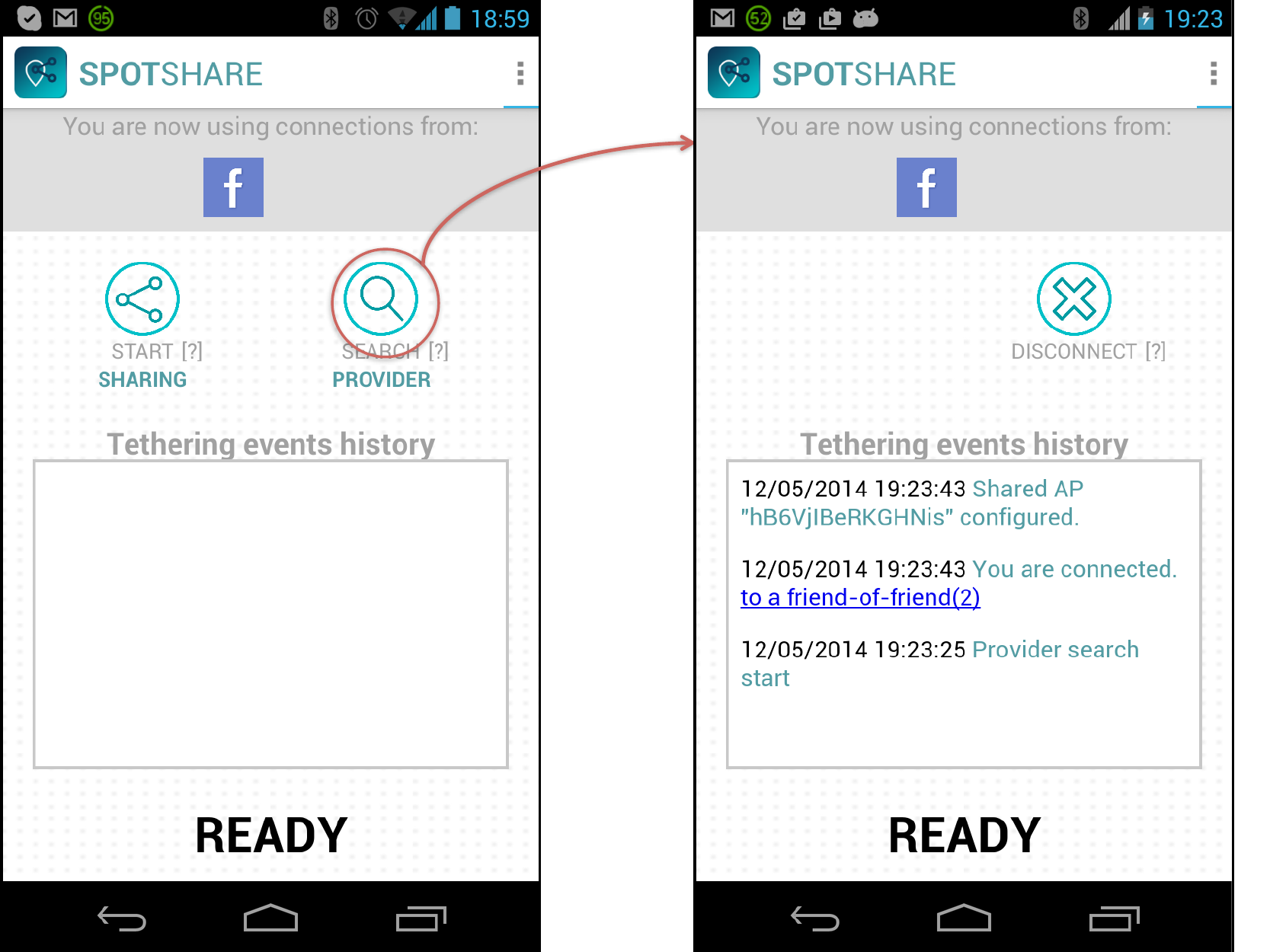}
 		\caption{\spotshare}
                 \label{subfig:spotshare}
        \end{subfigure}
	\qquad
	\begin{subfigure}[b]{\columnwidth}
		\centering 	
		\includegraphics[width=0.9\textwidth]{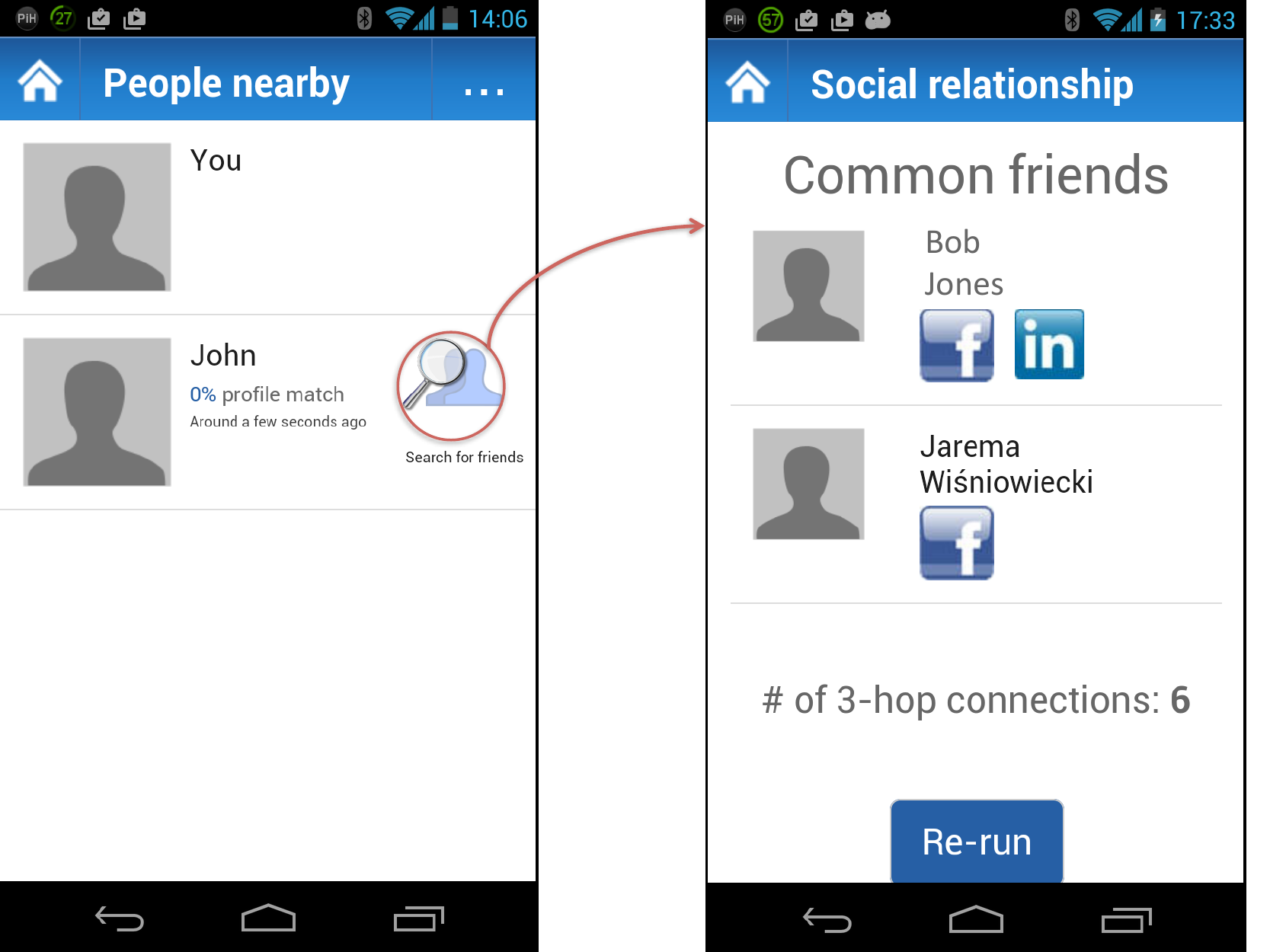}
 		\caption{\nearbypeople}
                 \label{subfig:nearbypeople}
         \end{subfigure}
        \vspace{-0.2cm}
\caption{Screenshots of the \spotshare and \nearbypeople apps.}
        \vspace{-0.2cm}
\label{fig:apps}
\end{figure*}
\fi

To illustrate \distest's relevance and practicality, we integrate it
into two Android apps, %
\spotshare and \nearbypeople, supporting both Facebook and LinkedIn.

\descr{\spotshare\footnote{\url{https://play.google.com/store/apps/details?id=org.sesy.tetheringapp}}} is an extension of
TetheringApp, presented in~\cite{bfpsi}. It allows a user to provide tethered Internet access to other \spotshare users %
(where access to the tethering hotspot is protected by a password) so that access control policies can be based on social relationships.
For instance, the user can decide to allow tethered access only to friends of friends: to this end, \spotshare uses \distest in order to determine, in a privacy-preserving way, if the specified social relationship holds. If so, the password is securely and automatically sent to the requesting device. %
In the current version of \spotshare, we do not enable discovery of social paths beyond two hops, as we assume that most users would not want to allow people with whom they have no common friends to tether off their smartphone, but removing this constraint is trivial.
\ifsubmission
\else
In Figure~\ref{subfig:spotshare}, we present two screenshots of the app.
\fi

\descr{\nearbypeople\footnote{\url{https://se-sy.org/projects/pet/nearbypeople.html}}} is a ``friend radar'' app
allowing users to interact with people around them and 
discovering common friends shared with users of nearby devices, as well as social path lengths, without having to broadcast their social profiles or rely on a central server. It relies on the privacy guarantees of \distest and the SCAMPI opportunistic router~\cite{scampi} for device-to-device communication.
A preliminary version of the app was successfully tested at the ACM CCS Workshop on Smart Energy Grid Security Workshop.
In Figure~\ref{subfig:nearbypeople}, we also show two screenshots.

\section{Related work}
\label{sec:related}

\descr{Privately discovering social relationships.}
\update{Nagy et al.~\cite{bfpsi} introduce \foffinder, reviewed in Section~\ref{subsec:system-model},
combining bearer capabilities with BFPSI/PSI-CA to allow OSN
users to discover, respectively, the identity or the number of their common friends in a private, authentic, and
decentralized way. While we build on the concept of capabilities and rely on BFPSI, recall that \foffinder suffers from an inherent bootstrapping problem and is limited to the discovery of social paths to OSN users that are two hops away.} Our work does not only address \foffinder's limitations via a novel methodology, but also  presents the full-blown implementation of a scalable server architecture and 
a modular Android client library enabling developers to easily integrate \distest into their applications.

Mezzour et al.~\cite{cans09} also describe techniques for decentralized
path discovery in social networks. 
They use a notion 
similar to capabilities to represent friendships
and hashing to derive higher-order capabilities, however, their scheme distributes a {\em different}
capability on behalf of a given user to every other user,
while \distest distributes the same capability to all users at a given
distance. \cite{cans09}'s computational/communication overhead is significantly higher than that of \distest: the former requires two PSI runs, with sets of size equal to the total number of paths from a node up to the maximum supported path length, whereas, the latter only requires a single BFPSI run, with input sets as big as the number of paths that have length equal to half the maximum supported path 
\ifsubmission
\else length. Table~\ref{tab:input-study} compares the expected number of input items for \distest and~\cite{cans09} based on the datasets introduced in Section~\ref{sec:evaluation}.
\fi
Furthermore,~\cite{cans09} incurs the same bootstrapping problem as~\cite{bfpsi}: if a friend \userA of user \user does not participate in the system, \user cannot detect paths to some other user \userB that go through \userA. 
Finally, \cite{cans09} aims to build a
decentralized social network, while we aim to
bootstrap the system based on existing centralized social networks. %

\ifsubmission
\else
\begin{table}[ttt]
	\vspace{0.15cm}
	\centering
	\resizebox{0.95\columnwidth}{!}{%
	\begin{tabular}{|c|c|r|r|}
	\hline
	\textbf{Dataset} & \textbf{Path length} & \textbf{Social Pal} & \textbf{Mezzour et al.~\cite{cans09}} \\ \hline
	\multirow{3}{*}{Social Filter} & 2 & 29.62 $\pm$ 40.12 & 1,008.12 $\pm$ 1,535.26 \\ \cline{2-4}
	& 3 & 1,037.74 $\pm$ 1,575.38 & 17,093.12 $\pm$ 20,159.86 \\ \cline{2-4}
	& 4 & 1,037.74 $\pm$ 1,575.38 & 125226 $\pm$ 96054.86 \\ \hline
	\multirow{3}{*}{MHRW} & 2 & 3.74 $\pm$ 2.63 & 205.87 $\pm$ 295.99 \\ \cline{2-4}
	& 3 & 209.61 $\pm$ 298.62 & 1,589.78 $\pm$ 2,212.04 \\ \cline{2-4}
	& 4 & 209.61 $\pm$ 298.62 & 40,798.78 $\pm$ 46,918.74 \\ \hline
	\multirow{3}{*}{BFS} & 2 & 55.38 $\pm$ 95.45 & 65,76.99 $\pm$ 8,606.57 \\ \cline{2-4}
	& 3 & 6,632.37 $\pm$ 8,702.02 & 154,059.99 $\pm$ 15,1526.57 \\ \cline{2-4}
	& 4 & 6,632.37 $\pm$ 8,702.02 & 1,255,889.99 $\pm$ 525,882.57 \\ \hline
	\end{tabular}	
	}
	        \vspace{-0.2cm}
	\caption{Average number of input items to PSI in \distest and Mezzour et al.~\cite{cans09} based on the datasets introduced in Section~\ref{sec:evaluation}.}
	        \vspace{-0.3cm}
	\label{tab:input-study}
\end{table}
\fi

Liao et al.~\cite{s-match} present a privacy-preserving social matching protocol based on property-preserving encryption (PPE), which however relies on a centralized approach.
Li et al.~\cite{Findu2011} then propose a set of protocols for privacy-preserving matching of attribute sets of different OSN users.
\ifsubmission
\else
Similar work include~\cite{DDQZ2011},~\cite{ZZSY2012}, and~\cite{ZL12}.
\fi
Private friend discovery has also been investigated in~\cite{HCE11}
and~\cite{VENETA}, which do not provide authenticity as they are vulnerable 
to malicious users claiming non-existent friendships.
While~\cite{DMP11} addresses the authenticity problem, it unfortunately comes at the cost of relying on relatively expensive
cryptographic techniques (specifically, a number of modular exponentiations linear
in the size of friend lists and a quadratic number of modular multiplications).

\ifsubmission
\else
    Smokescreen~\cite{cox2007smokescreen}, SMILE~\cite{manweiler2009smile},
    and PIKE~\cite{apolinarski2013pike} support secure/private device-to-device handshake
    and proximity-based communication.
\fi
Lentz et al.~introduce SDDR~\cite{sddr}, which allows a device
to establish a {\em secure encounter} -- i.e., a secret key -- with every device in short radio range,
and can be used to recognize previously encountered users, 
while providing strong unlinkability guarantees.
The EnCore platform~\cite{aditya-2014-encore}  builds on SDDR to provide
privacy-preserving interaction between nearby devices, as well as event-based communication
for mobile social applications. %

\descr{Building on Social Relationships.} Prior work has also focused on building services on top of existing social relationships.
Cici et al.~\cite{carpooling} use OSNs to assess the potential of ride-sharing services, showing that these would be very successful if users shared rides with friends of their friends. 
Sirivianos et al.~\cite{social-filter} propose a collaborative spam mitigation system
leveraging social networks of administrators, while~\cite{norcie2013bootstrapping}
and~\cite{sirivianos2012assessing} use OSNs to verify the veracity of online assertions.
\update{Freedman and Nicolosi~\cite{DBLP:conf/iptps/FreedmanN07} describe a system using social network for trust establishment in the context of email white-listing, by verifying the existence of common friends. Besides not discovering paths longer than two, \cite{DBLP:conf/iptps/FreedmanN07} also does not address the issue of friendships' authenticity -- unlike \distest.}

Daly et al.~\cite{DH07} present a routing protocol (called SimBet) for DTN networks based on social network data. Their protocol attempts to identify a routing bridge node based on the concept of centrality and transitivity of social networks. Li et al.~\cite{self-selfish-dtn-routing} design another DTN routing protocol (called Social Selfishness Aware Routing) which takes into account user's social selfishness and willingness to forward data only to nodes with sufficiently strong social ties. Other work~\cite{bubble-rap,vaad,roadcast} also propose adjusting message forwarding based on some social metrics. 

\descr{OSN Properties.} Another line of work has studied properties of OSNs. Ugander et al.~\cite{facebook-anatomy} and Backstrom et al.~\cite{backstrom2011} study the structure of Facebook social graph, revealing that the average social path length suggested by the ``small world experiment"~\cite{milgram1967small} (i.e., six) does not apply for Facebook, as the majority of people are separated by a 4-hop path. 
\ifsubmission
\else
    Gilbert et al.~\cite{gilbert-tie-strength-2} define the relationship between tie strengths (i.e., the importance of a social 
    relationship between two users) and various variables retrieved from the OSN social graph. 
    In~\cite{ego-networks-arnaboldi}, Arnaboldi et al.~investigate the link between 
    the tie strength definition (given by Granovotter~\cite{granovetter1973}) and a composition of factors describing the emotional 
    closeness in online relationships. They demonstrate the existence of the \textit{Dunbar number} (i.e., the maximum number of 
    people a user can actively interact with) for Facebook. 
    In follow-up work~\cite{ego-facebook,ego-twitter}, they also show the existence of four hierarchical layers of social 
    relationships inside ego networks. 
    Existence of the Dunbar number is also shown for Twitter in~\cite{gonalves2011modeling}. 
    Finally, Saram\"{a}ki et al.~\cite{social-signatures} find an uneven distribution of tie strengths within ego networks that is 
    characterized by the presence of a few strong and a majority of weak ties.
\fi

\section{Conclusion}
\label{sec:conclusions}

This paper presented \distest\ -- a system geared to privately
estimate the social path length between two 
social network users.
We demonstrated its effectiveness both 
analytically and empirically, showing that, for any two OSN users,
\distest discovers all social paths of length two and a significant
portion of longer paths. Using different samples of the Facebook graph, 
we showed that even when only 20\% of the OSN users use the system, we discover more than 40\% of all
paths between any two users, and 70\% with 40\% of users.

We also implemented a scalable server-side architecture and a modular client library bringing \distest to the real world. Our deployment supports Facebook and LinkedIn integration and allows developers to easily incorporate it in their projects. \distest can be used in
a number of applications %
where, by relying on
the (privacy-preserving) estimation of social path length, users
can make informed trust and access control decisions.

In future work, we will augment \distest with information about
the 
{\em tie strength}~\cite{ego-networks-arnaboldi,gilbert-tie-strength-2}
between users, and present a usability study of some sample applications
built on top of the system.

\descr{Acknowledgments.} We thank Minas Gjoka and Michael Sirivianos for sharing the Facebook datasets, Swapnil Udar for helping with the \spotshare implementation, and Jussi Kangasharju, Pasi Sarolahti, Cecilia Mascolo, Panos Papadimitratos, and Narges Yousefnezhad for providing feedback on the paper. Simon Eberz suggested the idea of signed Bloom Filters discussed in Section~\ref{subsec:privacy-consideration}. This work was partially supported by the Academy of Finland's ``Contextual Security'' project (274951), the EC FP7 PRECIOUS project (611366), and the EIT ICT Labs.

\end{document}